\documentclass[12pt,english]{article}
\usepackage{mathptmx}

\usepackage[LGR,T1]{fontenc}
\usepackage[latin9]{inputenc}
\usepackage{geometry}
\usepackage{fancyhdr}
\usepackage{amsmath}
\usepackage{amssymb}
\usepackage{cancel}
\usepackage{setspace}
\usepackage{esint}
\makeatletter

\DeclareRobustCommand{\greektext}{%
  \fontencoding{LGR}\selectfont\def\encodingdefault{LGR}}
\DeclareRobustCommand{\textgreek}[1]{\leavevmode{\greektext #1}}
\ProvideTextCommand{\~}{LGR}[1]{\char126#1}

\newcommand{\lyxaddress}[1]{
	\par {\raggedright #1
	\vspace{1.4em}
	\noindent\par}
}
\newenvironment{lyxcode}
	{\par\begin{list}{}{
		\setlength{\rightmargin}{\leftmargin}
		\setlength{\listparindent}{0pt}
		\raggedright
		\setlength{\itemsep}{0pt}
		\setlength{\parsep}{0pt}
		\normalfont\ttfamily}%
	 \item[]}
	{\end{list}}

\date{}
\numberwithin{equation}{section}

\makeatother

\usepackage{babel}
\begin{document}
\global\long\def\bra#1{\left\langle #1\right|}
\global\long\def\ket#1{\left|#1\right\rangle }
\global\long\def\braket#1#2{\left\langle #1|#2\right\rangle }
\global\long\def\ketbra#1#2{\left|#1\right\rangle \left\langle #2\right|}

{\tiny{}}\global\long\def\b#1{\Bra{#1}}
{\tiny{}}\global\long\def\k#1{\Ket{#1}}
{\tiny{}}\global\long\def\set#1{\Set{#1}}
{\tiny{}}\global\long\def\bk#1{\Braket{#1}}
{\tiny{}}\global\long\def\norm#1{\left\Vert #1\right\Vert }
{\tiny\par}

\global\long\def\broket#1#2#3{\bra{#1}#2\ket{#3}}

\global\long\def\clop#1{\left[#1\right)}
\global\long\def\opcl#1{\left(#1\right]}

\global\long\def\beC#1#2{\begC#2{#1}}
\global\long\def\enC#1#2{\endC#2{#1}}

{\tiny{}}\global\long\def\at#1{\left.#1\right|}
{\tiny\par}

\global\long\def\paren#1{\left(#1\right.}
\global\long\def\thesis#1{\left.#1\right)}

\global\long\def\del{\nabla}
\global\long\def\cross{\times}
\global\long\def\div{\del\cdot}
\global\long\def\rot{\del\cross}
{\tiny{}}\global\long\def\lap{\nabla^{2}}
{\tiny{} }{\tiny\par}

\global\long\def\divv{\vec{\nabla}\cdot}
\global\long\def\delv{\vec{\nabla}}
\global\long\def\rotv{\vec{\nabla}\cross}

\global\long\def\diff#1#2{\frac{\partial#1}{\partial#2}}
\global\long\def\Diff#1#2{\frac{d#1}{d#2}}

\global\long\def\up{\uparrow}
\global\long\def\down{\downarrow}
\global\long\def\tg{\prime}
\global\long\def\dag{\dagger}

\global\long\def\sinc{\mbox{sinc}}
\global\long\def\Tr{\mbox{Tr}}
\global\long\def\const{\mbox{const}}
{\tiny{}}\global\long\def\trace{{\rm trace}}
{\tiny\par}

\global\long\def\defined{\equiv}
\global\long\def\r{\rightarrow}

\global\long\def\matt#1#2#3#4{\left(\begin{array}{cc}
 #1  &  #2\\
 #3  &  #4 
\end{array}\right)}
\global\long\def\colt#1#2{\left(\begin{array}{c}
 #1\\
 #2 
\end{array}\right)}
\global\long\def\rowt#1#2{\left(\begin{array}{cc}
 #1  &  #2\end{array}\right)}

\global\long\def\identity{\mathbf{1}}
\global\long\def\paulix{\begin{pmatrix}  &  1\\
 1 
\end{pmatrix}}
\global\long\def\pauliy{\begin{pmatrix}  &  -i\\
 i 
\end{pmatrix}}
\global\long\def\pauliz{\begin{pmatrix}1\\
  &  -1 
\end{pmatrix}}

{\tiny{}}\global\long\def\vecpro#1#2#3#4#5#6{\left|\begin{array}{ccc}
 \hat{x}  &  \hat{y}  &  \hat{z}\\
 #1  &  #2  &  #3\\
 #4  &  #5  &  #6 
\end{array}\right|}
{\tiny\par}

\global\long\def\undercom#1#2{\underset{_{#2}}{\underbrace{#1}}}
{\tiny{}}\global\long\def\explain#1#2{\underset{\mathclap{\overset{\uparrow}{#2}}}{#1}}
{\tiny{}}\global\long\def\iexplain#1#2{\underset{\mathclap{\overset{\big{\uparrow}}{#2}}}{#1}}
{\tiny{}}\global\long\def\bracedown#1#2{\underbrace{#1}_{\mathclap{#2}}}
{\tiny\par}

{\tiny{}}\global\long\def\explainup#1#2{\overset{\mathclap{\underset{\downarrow}{#2}}}{#1}}
{\tiny{}}\global\long\def\iexplainup#1#2{\overset{\mathclap{\underset{\big{\downarrow}}{#2}}}{#1}}
{\tiny{}}\global\long\def\braceup#1#2{\overbrace{#1}^{\mathclap{#2}}}
{\tiny\par}

{\tiny{}}\global\long\def\rmto#1#2{\cancelto{#2}{#1}}
{\tiny{}}\global\long\def\rmpart#1{\cancel{#1}}
\global\long\def\conv#1#2{\underset{_{#1\rightarrow#2}}{\longrightarrow}}

\global\long\def\MA{\mathcal{A}}
\global\long\def\MB{\mathcal{B}}
\global\long\def\MC{\mathcal{C}}
\global\long\def\MD{\mathcal{D}}
\global\long\def\ME{\mathcal{E}}
\global\long\def\MF{\mathcal{F}}
\global\long\def\MG{\mathcal{G}}
\global\long\def\MH{\mathcal{H}}
\global\long\def\MI{\mathcal{I}}
\global\long\def\MJ{\mathcal{J}}
\global\long\def\MK{\mathcal{K}}
\global\long\def\ML{\mathcal{L}}
\global\long\def\MM{\mathcal{M}}
\global\long\def\MN{\mathcal{N}}
\global\long\def\MO{\mathcal{O}}
\global\long\def\MP{\mathcal{P}}
\global\long\def\MQ{\mathcal{Q}}
\global\long\def\MR{\mathcal{R}}
\global\long\def\MS{\mathcal{S}}
\global\long\def\MT{\mathcal{T}}
\global\long\def\MU{\mathcal{U}}
\global\long\def\MV{\mathcal{V}}
\global\long\def\MW{\mathcal{W}}
\global\long\def\MX{\mathcal{X}}
\global\long\def\MY{\mathcal{Y}}
\global\long\def\MZ{\mathcal{Z}}
\global\long\def\BA{\mathbb{A}}
\global\long\def\BB{\mathbb{B}}
\global\long\def\BC{\mathbb{C}}
\global\long\def\BD{\mathbb{D}}
\global\long\def\BE{\mathbb{E}}
\global\long\def\BF{\mathbb{F}}
\global\long\def\BG{\mathbb{G}}
\global\long\def\BH{\mathbb{H}}
\global\long\def\BI{\mathbb{I}}
\global\long\def\BJ{\mathbb{J}}
\global\long\def\BK{\mathbb{K}}
\global\long\def\BL{\mathbb{L}}
\global\long\def\BM{\mathbb{M}}
\global\long\def\BN{\mathbb{N}}
\global\long\def\BO{\mathbb{O}}
\global\long\def\BP{\mathbb{P}}
\global\long\def\BQ{\mathbb{Q}}
\global\long\def\BR{\mathbb{R}}
\global\long\def\BS{\mathbb{S}}
\global\long\def\BT{\mathbb{T}}
\global\long\def\BU{\mathbb{U}}
\global\long\def\BV{\mathbb{V}}
\global\long\def\BW{\mathbb{W}}
\global\long\def\BX{\mathbb{X}}
\global\long\def\BY{\mathbb{Y}}
\global\long\def\BZ{\mathbb{Z}}

{\tiny{}}\global\long\def\eps{\varepsilon}
\global\long\def\ph{\varphi}
\global\long\def\th{\theta}
\global\long\def\d{\partial}
\global\long\def\h{\hbar}
{\tiny{}}\global\long\def\angstrom{\overset{\circ}{\mathrm{A}}}
{\tiny\par}

{\tiny{}}\global\long\def\dbar{{\mathchar'26\mkern-12mu  {\rm d} }}
{\tiny\par}

\global\long\def\ind#1#2#3{#1_{\hphantom{#2}#3}^{#2}}
\global\long\def\rind#1#2#3{#1_{#2}^{\hphantom{#2}#3}}

\global\long\def\mom#1#2{\momentum{#1}{#2}}
\global\long\def\momb#1#2{\momentum[bot]{#1}{#2}}

\global\long\def\vlup#1{\vertexlabel^{{#1}}}
\global\long\def\vldown#1{\vertexlabel_{{#1}}}

{\tiny{}}\global\long\def\wcal#1#2#3{\contraction{}{#1}{#2}{#3}#1#2#3}
{\tiny{}}\global\long\def\wcah#1#2#3{\contraction[2ex]{}{#1}{#2}{#3}#1#2#3}
{\tiny\par}

{\tiny{}}\global\long\def\wcbl#1#2#3{\bcontraction{}{#1}{#2}{#3}#1#2#3}
{\tiny{}}\global\long\def\wcbh#1#2#3{\bcontraction[2ex]{}{#1}{#2}{#3}#1#2#3}
{\tiny\par}

{\tiny{}}\global\long\def\doublewick#1#2#3#4#5#6#7{\contraction{}{#1}{#2#3#4}{#5}\contraction[2ex]{#1#2}{#3}{#4#5#6}{#7}#1#2#3#4#5#6#7}
{\tiny\par}

{\tiny{}}\global\long\def\vec#1{\mathbf{#1}}
{\tiny{}}\global\long\def\hatt#1#2{\hat{#1}_{#2}}
{\tiny\par}

\global\long\def\limm#1#2{\lim\limits _{#1\rightarrow#2}}

\global\long\def\sig#1#2{\sum\limits _{#1}^{#2}}
\global\long\def\intt#1#2{\int\limits _{#1}^{#2}}
 {\tiny{}}\global\long\def\inftop{\intop_{-\infty}^{\infty}}
{\tiny\par}

{\tiny{}}\global\long\def\ra{\quad\Rightarrow\quad}
{\tiny{}}\global\long\def\sp{\hfill}
{\tiny{}}\global\long\def\comm{\quad,\quad}
{\tiny\par}

{\tiny{}}\global\long\def\e#1{\cdot10^{#1}}
{\tiny\par}

\global\long\def\ofr{\left(\overrightarrow{r}\right)}
\global\long\def\kb{k_{B}}
\global\long\def\kbt{k_{B}T}

\title{Understanding Higher-Spin Gravity through Bilocal Holography for
Free Fermions}

\author{Tomer Solberg}
\maketitle

\lyxaddress{Department of Particle Physics and Astrophysics, Weizmann Institute
of Science, Rehovot 76100, Israel}
\begin{lyxcode}
E-mail:~Tomer.Solberg@gmail.com
\end{lyxcode}
\begin{abstract}
We consider a simple theory of $N$ free fermions in $d$ dimensions
with $O\left(N\right)$ or $U\left(N\right)$ symmetry. The singlet
sector of this theory is expected from holography to be dual to the
notoriously complicated Vasiliev gravity. By defining bilocal fields
we find an explicit holographic mapping between the two theories,
similar to what was done previously for scalars, which allows us to
construct a quantum version of Vasiliev gravity, including loop ($=1/N$)
corrections. Further more, such a mapping establishes a non-supersymmetric
instance of the AdS/CFT duality.
\end{abstract}

\section{Introduction}

It has long been known that some $d$-dimensional quantum field theories
are dual to theories of gravity in $\left(d+1\right)$-dimensions
\cite{key-1}. This kind of duality is known as holography, and its
most known and studied example is the AdS/CFT correspondence, in which
the gravitational theory (for example a string theory) is living in
anti-de Sitter (AdS) spacetime, and it is dual to a conformal field
theory on a space that is conformally equivalent to the boundary of
AdS \cite{key-4}. This is especially useful in the case where the
field theory has some gauge group with a fundamental representation
of size $N$, where $N$ is taken to be large. In the large $N$ limit,
the dual gravitational theory is then weakly coupled, with a coupling
proportional to $1/N$ \cite{key-3}. Generally speaking, the AdS/CFT
correspondence is a weak/strong duality. If we focus on the large
$N$ limit then the gravitational theories are weakly coupled. However,
these gravitational theories can look approximately local, in the
sense that higher derivative corrections to their interactions are
small at the scale of the curvature of space-time, only when the field
theories are strongly coupled (in the sense of having a large \textquoteleft t
Hooft coupling). This is the regime where most of the research on
holography is focused, since it can be used to relate standard, almost
classical, gravitational theories, to strongly coupled field theories
(some of which are interesting for applications like the study of
QCD).

In this work we study the opposite regime -- the extreme limit where
we have a free field theory on the field theory side of the equivalence.
Somewhat surprisingly, such theories are still equivalent (in an appropriate
sense) to gravitational theories, which can be weakly coupled if the
free field theory has a large number of degrees of freedom. The first
such example of the AdS/CFT duality was suggested by Klebanov and
Polyakov \cite{key-5}, who looked at the simple theory of $N$ real
free fields $\ML=\frac{1}{2}\left(\d\phi_{i}\right)^{2}$ which has
an $O\left(N\right)$ symmetry, and infinitely many conserved currents
which are $O\left(N\right)$ singlets 
\begin{equation}
J_{\left(\mu_{1}\cdots\mu_{s}\right)}=\phi_{i}\left(\overleftrightarrow{\d}_{\mu_{1}}\cdots\overleftrightarrow{\d}_{\mu_{s}}\right)\phi_{i},
\end{equation}
and observed that these are in one-to-one correspondence with with
the spectrum of massless higher-spin fields in the minimal bosonic
theory in AdS, which is known as Vasiliev gravity \cite{key-8}. The
latter is a generalization of Einstein's gravity in AdS space, which
includes a tower of massless degrees of freedom of spins $s=1,2,3,4,\dots$
(for $O\left(N\right)$ models only even spins appear, while all integer
spins appear for $U\left(N\right)$ models), and is considered to
be immensely complicated. In this paper we discuss similarly the fermionic
case, first studied by Sezgin and Sundell \cite{key-13}. We continue
the work by Aharony, Chester and Urbach \cite{key-9}, which have
explicitly found this mapping for scalar theories, and generalize
it from free scalars to free fermions. This case is more complicated
because of the extra spinor indices of the fermions.

\subsection{The story with scalars so far }

We begin by reviewing the explicit mapping for the scalar case \cite{key-9}.
Our goal is to try and translate explicitly from the CFT side to the
bulk in order to build a quantum model of Vasiliev gravity. To achieve
this, we follow a recent proposal made by Jevicki et al. \cite{key-6}
for a way to analyze this duality using bilocal fields. The bilocal
fields are 
\begin{equation}
G\left(x_{1},x_{2}\right)=\frac{1}{N}\sum_{i}\phi_{i}\left(x_{1}\right)\phi_{i}\left(x_{2}\right)
\end{equation}
and they contain all information on $O\left(N\right)$-invariant operators,
including the local operators $\MO_{s}=\phi_{i}\undercom{\d\cdots\d}s\phi_{i}$
which appear in the expansion of $G$ as $x_{2}\r x_{1}$. We can
rewrite the partition function in terms of the bilocal fields 
\begin{equation}
\intop\MD\phi_{i}e^{-\intop d^{d}x\frac{1}{2}\left(\d\phi_{i}\right)^{2}}\r\intop\MD Ge^{-S\left(G\right)}.
\end{equation}

To achieve this, we discretize space-time to $V=M^{d}$ sites, such
that $\phi_{i}\left(x\right)$ is a length $V$ vector and $G\left(x_{1},x_{2}\right)$
is a $V\times V$ matrix. We work in the large $N$ limit in which
we also assume $N\geq V$. We introduce the bilocal field matrix multiplication
which is defined by 
\begin{equation}
\left(GH\right)\left(x_{1},x_{2}\right)=\intop d^{d}x_{3}G\left(x_{1},x_{3}\right)H\left(x_{3},x_{2}\right).
\end{equation}
We then manipulate the path integral in the following way 
\begin{align}
\int\MD\phi_{i}e^{-S_{0}\left(G\left(\phi\right)\right)} & =\int\MD G\MD\phi_{i}e^{-S_{0}\left(G\right)}\delta\left(G-\frac{1}{N}\sum_{i}\phi_{i}\phi_{i}\right)\nonumber \\
 & =\int\MD Ge^{-S_{0}\left(G\right)}\int\MD\Sigma\MD\phi_{i}e^{\Tr\left(i\frac{N}{2}\Sigma G-\frac{i}{2}\phi_{i}\Sigma\phi_{i}\right)}\nonumber \\
 & =\int\MD Ge^{-S_{0}\left(G\right)}\int\MD\Sigma\left(\det\Sigma\right)^{-N}e^{\Tr\left(i\frac{N}{2}\Sigma G\right)}
\end{align}

where we have introduced another bilocal field $\Sigma$. The integral
over $\Sigma$ can be carried out by a change of variables $\tilde{\Sigma}=\Sigma G$
so that $\MD\Sigma=\left(\det G\right)^{-V}\MD\tilde{\Sigma}$, and
so up to a constant function of $N,V$
\begin{equation}
\int\MD\phi_{i}e^{-S_{0}\left(G\left(\phi\right)\right)}=C_{N}\int\MD Ge^{-S_{0}\left(G\right)}\left(\det G\right)^{N-V}.
\end{equation}
Plugging in $S_{0}\left(G\right)$ which is just the original action
of $\phi_{i}$ expressed in terms of $G$, we get
\begin{equation}
S\left(G\right)=\frac{1}{2}\intop d^{d}x\left[-N\left(\d_{x}^{2}G\left(x,x^{\tg}\right)\right)_{x^{\tg}=x}\right]-\left(N-V\right)\Tr\log G.\label{eq:action}
\end{equation}

One should note that $V$ is an arbitrary regulator. For other regulators
$V$ will be replaced by an appropriate cut-off dependent constant,
which can be viewed as a one-loop renormalization of an action that
is proportional to $N$. Thus, the loop expansion of (\ref{eq:action})
is an expansion in powers of $1/N$. This action can be expanded around
a simple saddle point 
\begin{align}
G_{0}\left(x_{1},x_{2}\right) & =\frac{1}{N}\sum_{i}\left\langle \phi_{i}\left(x_{1}\right)\phi_{i}\left(x_{2}\right)\right\rangle ,\\
G & =G_{0}+\frac{1}{\sqrt{N}}\eta,
\end{align}
so that we can write the action for $\eta$. This action will have
a Gaussian part as well as perturbation terms of the form $\eta^{n}$
for any $n>2$ due to the $\Tr\log G$ term in $S\left[G\right]$.
This will give us an intricate perturbation theory in the parameter
$1/N$ which we can work out. Our advantage is that we know what the
final results must be, as this is equivalent to the free theory which
we can work out to any order. For example in $d=3$ we know that $G_{0}$
and the two point function in $\eta$ are exactly (denoting $x_{ij}=x_{i}-x_{j}$)
\begin{align}
G_{0}\left(x_{1},x_{2}\right) & \sim\frac{1}{\left|x_{12}\right|}\\
\left\langle \eta\left(x_{1},x_{2}\right)\eta\left(x_{3},x_{4}\right)\right\rangle  & \sim\frac{1}{\left|x_{13}\right|\left|x_{24}\right|}+\frac{1}{\left|x_{14}\right|\left|x_{23}\right|}\label{eq:corr}
\end{align}
with no further dependence on $N$.

This means that all the loop diagrams in the $\eta$ theory have to
cancel out, which has indeed been shown. The $\eta$ field can then
be expanded in the basis of the conformal 3-pt. functions 
\begin{align}
\left\langle \phi\left(x_{1}\right)\phi\left(x_{2}\right)\MO_{\Delta}^{\mu_{1}\dots\mu_{J}}\left(x_{3}\right)\right\rangle  & =\frac{Z^{\mu_{1}}\cdots Z^{\mu_{J}}-\text{traces}}{\left|x_{12}\right|^{2\Delta_{0}-\Delta}\left|x_{13}\right|^{\Delta}\left|x_{23}\right|^{\Delta}}\\
Z^{\mu} & =\frac{\left|x_{13}\right|\left|x_{23}\right|}{\left|x_{12}\right|}\left(\frac{x_{13}^{\mu}}{x_{13}^{2}}-\frac{x_{23}^{\mu}}{x_{23}^{2}}\right)
\end{align}
where $\Delta_{0}=\frac{d-2}{2}$ is the conformal weight of a free
field $\phi$. The expansion is \cite{key-10}
\begin{equation}
\eta\left(x_{1},x_{2}\right)=\sum_{J=0}^{\infty}\intop_{C_{J}}\frac{d\Delta}{2\pi i}\intop\frac{d^{d}y}{J!\left(\frac{d}{2}-1\right)_{J}}c_{\Delta,J}^{\mu_{1}\dots\mu_{J}}\left(y\right)\left\langle \phi\left(x_{1}\right)\phi\left(x_{2}\right)\MO_{\mu_{1}\dots\mu_{J}}^{\Delta,J}\left(y\right)\right\rangle 
\end{equation}

where the $c_{\Delta,J}^{\mu_{1}\dots\mu_{J}}\left(y\right)$ are
the coefficients of the decomposition, and the contour $C_{J}$ in
the $\Delta$ integral runs over the principal series $\text{Re}(\Delta)=d/2$,
as well as around possible poles in the discrete series when $d<4$.
Using this expansion, we can write the quadratic action using the
action of the bilocal Laplacian (which appears in the quadratic action
for $\eta$), 

\begin{equation}
\del_{1}^{2}\del_{2}^{2}\left\langle \phi\left(x_{1}\right)\phi\left(x_{2}\right)\MO_{\Delta,J}\left(y\right)\right\rangle =\lambda_{\Delta,J}\left\langle \tilde{\phi}\left(x_{1}\right)\tilde{\phi}\left(x_{2}\right)\MO_{\Delta,J}\left(y\right)\right\rangle ,\label{eq:biloclap}
\end{equation}
where $\tilde{\phi}$ denotes the shadow transform of $\phi$ with
conformal weight $\tilde{\Delta}_{0}\equiv d-\Delta_{0}$, and 
\begin{align}
\lambda_{\Delta,J} & =\left(M_{\Delta,J}^{2}-M_{d+J,J}^{2}\right)\left(M_{\Delta,J}^{2}-M_{d+J-2,J}^{2}\right),\\
M_{\Delta,J}^{2} & =\Delta\left(\Delta-d\right)-J.
\end{align}

One can write a similar expansion of transverse traceless bulk fields
living on a fixed AdS background $\Phi_{J}\left(X,S\right)$, where
$X_{a}$ is a bulk coordinate and $S^{a}$ is an auxiliary null vector
variable keeping track of spin indices (this notation is explained
in detail in appendix A), using the bulk-to-boundary propagators of
massive fields on $AdS$ with spin $J$ 
\begin{equation}
G_{\Delta,J}\left(X_{1},X_{2};S_{1},S_{2}\right)=\frac{\left(X_{12}\left(S_{1}\cdot S_{2}\right)+2\left(S_{1}\cdot X_{2}\right)\left(S_{2}\cdot X_{1}\right)\right)^{J}}{X_{12}^{\Delta+J}}.\label{eq:bulk2bdry}
\end{equation}
Here $\left(X_{2},S_{2}\right)$ is a coordinate on the boundary,
given in bulk notation, and $X_{12}=-2X_{1}\cdot X_{2}$. These are
unique solutions to the $AdS$ Laplace equation 
\begin{equation}
\del_{X_{1}}^{2}G_{\Delta,J}\left(X_{1},X_{2};S_{1},S_{2}\right)=M_{\Delta,J}^{2}G_{\Delta,J}\left(X_{1},X_{2};S_{1},S_{2}\right),\label{eq:lapads}
\end{equation}
which satisfy the transversality condition 
\begin{equation}
\del_{X_{1}}\cdot K_{S_{1}}G_{\Delta,J}\left(X_{1},X_{2};S_{1},S_{2}\right)=0,
\end{equation}
where
\begin{align}
K_{S,a} & =\left(\frac{d-1}{2}+S\cdot\frac{\d}{\d S}\right)\frac{\d}{\d S^{a}},\label{eq:KS}
\end{align}
and the boundary condition (as $\left(X_{1},S_{1}\right)$ approaches
the boundary)
\begin{equation}
G_{\Delta,J}\left(\frac{1}{z}X_{1},X_{2};\frac{1}{z}S_{1},S_{2}\right)=z^{\Delta-J}\left\langle \MO\left(X_{2},S_{2}\right)\MO\left(X_{1},S_{1}\right)\right\rangle +z^{d-\Delta-J}S_{B}^{\Delta,J}\delta\left(X_{1},X_{2}\right)\left(S_{1}\cdot S_{2}\right)^{J}+\dots
\end{equation}

Here we used an AdS metric $ds^{2}=\left(dx_{\mu}^{2}+dz^{2}\right)/z^{2}$
and denoted the bulk shadow coefficient 
\begin{equation}
S_{B}^{\Delta,J}=\frac{\pi^{d/2}\Gamma\left(\Delta-d/2\right)}{\left(J+\Delta-1\right)\Gamma\left(\Delta-1\right)}.\label{eq:bulkshadow}
\end{equation}
The expansion is given by 
\begin{equation}
\Phi_{J}\left(X,S\right)=\intop_{P.S.}\frac{d\Delta}{2\pi i}\intop\frac{dY}{J!\left(\frac{d}{2}-1\right)_{J}}c_{\Delta,J}^{bulk}\left(Y,D_{Z}\right)G_{\Delta,J}\left(X,Y;S,Z\right).
\end{equation}
By writing the CFT coefficients $c_{\Delta,J}^{\mu_{1}\dots\mu_{J}}\left(x\right)$
in embedding space coordinates $c_{\Delta,J}\left(X,S\right)$, we
can find an explicit mapping between the boundary and bulk theories
through 
\begin{equation}
c_{\Delta,J}^{bulk}\left(X,S\right)=f_{\Delta,J}c_{\Delta,J}\left(X,S\right)
\end{equation}
for some arbitrary (up to certain consistency constraints) normalization
$f_{\Delta,J}$. Plugging this in gives us a linear mapping between
the bilocal and bulk fields 
\begin{align}
\Phi_{J}\left(X,W\right) & =\intop d^{d}x_{1}d^{d}x_{2}\MM_{J}\left(X,W;x_{1},x_{2}\right)\eta\left(x_{1},x_{2}\right)\\
\eta\left(x_{1},x_{2}\right) & =\sum_{J}\intop dX\MM_{J}^{-1}\left(x_{1},x_{2};X,K_{W}\right)\Phi_{J}\left(X,W\right),
\end{align}
The bulk quadratic action is then (for a specific choice of $f_{\Delta,J}$)
\begin{equation}
S^{\left(2\right)}\left[\Phi_{J}\right]=\sum_{J}\frac{\pi^{d/2}\Gamma\left(J+1\right)}{2^{J}\Gamma\left(J+\frac{d}{2}\right)}\intop\frac{dX}{\left(\frac{d-1}{2}\right)_{J}J!}\Phi_{J}\left(X,K_{W}\right)\left(\del_{X}^{2}-M_{d+J-2,J}^{2}\right)\left(\del_{X}^{2}-M_{d+J,J}^{2}\right)\Phi_{J}\left(X,W\right),
\end{equation}

and indeed the full bilocal action can be translated into a full (non-local)
bulk action which is an explicit quantum generalization of Vasiliev
gravity.

\section{Bi-local holography with fermions}

\subsection{Bilocal fermion action}

We start in $d\geq2$ dimensions, with $N$ fermion fields $\psi_{\alpha}^{i}$
with $i=1,\dots,N$ and $\alpha$ is a spinor index (assumed to be
Dirac for now). We have a global $U\left(N\right)$ symmetry, and
all singlets are given by the bilocal fields 
\begin{equation}
\sigma_{\alpha\beta}\left(x_{1},x_{2}\right)=\frac{1}{N}\overline{\psi}_{\alpha}^{i}\left(x_{1}\right)\psi_{\beta}^{i}\left(x_{2}\right).
\end{equation}

Note that for Majorana fermions we would have $O\left(N\right)$ symmetry.
The free fermion Lagrangian can be expressed in terms of $\sigma$
as 
\begin{equation}
\overline{\psi}_{\alpha}^{i}\gamma_{\alpha\beta}^{\mu}\d_{\mu}\psi_{\beta}^{i}\left(x\right)=\at{N\gamma_{\alpha\beta}^{\mu}\d_{\mu}^{\tg}\sigma_{\alpha\beta}\left(x,x^{\tg}\right)}_{x^{\tg}\r x}=\at{N\text{tr}\left(\cancel{\d}\sigma\right)}_{x^{\tg}\r x}
\end{equation}

where the trace is taken over spinor indices. The path integral is
\begin{align}
Z & =\intop\MD\overline{\psi}\MD\psi e^{-S\left(\overline{\psi},\psi\right)}\\
 & =\intop\MD\sigma\MD\overline{\psi}\MD\psi e^{-S_{0}\left(\sigma\right)}\prod_{x_{1},x_{2},\alpha,\beta}\delta\left(\sigma_{\alpha\beta}\left(x_{1},x_{2}\right)-\frac{1}{N}\overline{\psi}_{\alpha}^{i}\left(x_{1}\right)\psi_{\beta}^{i}\left(x_{2}\right)\right)\\
 & =\intop\MD\sigma e^{-S_{0}\left(\sigma\right)}\intop\MD\Sigma\MD\overline{\psi}\MD\psi e^{i\Tr\left(\Sigma\sigma-\frac{1}{N}\overline{\psi}^{i}\Sigma\psi^{i}\right)}\\
 & =\intop\MD\sigma e^{-S_{0}\left(\sigma\right)}\intop\MD\Sigma\left(\det\Sigma\right)^{N}e^{i\Tr\left(\Sigma\sigma\right)},
\end{align}
where $\Sigma$ is now a matrix both in position space and in spinor
space, and the trace is taken over both types of indices. Now define
$\tilde{\Sigma}=\Sigma\sigma$ so that $\MD\Sigma=\MD\tilde{\Sigma}\left|\text{det}\sigma\right|^{-K}$
with $K$ the number of (spacetime+spinor) DOF of $\sigma$. Integrating
over $\tilde{\Sigma}$ we get 
\begin{equation}
Z=\#\intop\MD\sigma e^{-S_{0}\left(\sigma\right)}\left|\det\sigma\right|^{-\left(N+K\right)}=\#\intop\MD\sigma e^{-S_{0}\left(\sigma\right)-\left(N+K\right)\Tr\log\sigma}
\end{equation}
where the Trace is taken over both spacetime and spinor indices. We
will ignore the $K$ factor as it can be removed with proper counter
terms as we did in the bosonic case. We can now expand $\sigma=\sigma_{0}+\rho/\sqrt{N}$
where $\sigma_{0}=\left\langle \overline{\psi}^{i}\psi^{i}\right\rangle /N$
so that we get $S\left(\rho\right)=S_{quad}+S_{int}$ where the quadratic
action is
\begin{equation}
S_{quad}=\frac{1}{2}\Tr\left(\sigma_{0}^{-1}\rho\right)^{2}=\frac{1}{2}\Tr\left(\cancel{\d}\rho\right)^{2}=-\frac{1}{2}\intop d^{d}x_{1}d^{d}x_{2}\rho_{\gamma\delta}\left(x_{1},x_{2}\right)\cancel{\d}_{1\beta}^{\gamma}\cancel{\d}_{2\alpha}^{\delta}\rho^{\alpha\beta}\left(x_{1},x_{2}\right).\label{eq:quadaction}
\end{equation}

The 2-pt. correlation function of the free fermions is given by 
\begin{align}
\left\langle \psi\left(x\right)\overline{\psi}\left(0\right)\right\rangle _{\alpha\beta} & =\intop\frac{d^{d}p}{\left(2\pi\right)^{d}}\frac{i\cancel{p}_{\alpha\beta}}{p^{2}}e^{-ip\cdot x}=-\cancel{\d}\intop\frac{d^{d}p}{\left(2\pi\right)^{d}}\frac{1}{p^{2}}e^{-ip\cdot x}\\
 & =\cancel{\d}i\frac{\Gamma\left(\frac{d}{2}-1\right)}{4\pi^{d/2}}\left(x^{2}\right)^{-\frac{d-2}{2}}=-i\gamma_{\alpha\beta}^{\mu}\frac{\Gamma\left(\frac{d}{2}\right)}{2\pi^{d/2}}x_{\mu}\left(x^{2}\right)^{-d/2}.
\end{align}

In order to make the conformal symmetry explicit at the level of the
path integral, we would like to use a basis that diagonalizes the
Cartan subalgebra of the conformal algebra. In a fermion mean field
theory (MFT), $\rho\left(x_{1},x_{2}\right)$ is in the tensor product
of two representations with a fermion primary of dimension $\Delta_{\psi}$
(for free fermions $\Delta_{\psi}=\frac{d-1}{2}$). Because we want
to decompose it into irreducible representations of the conformal
group, the basis is by definition the Clebsch-Gordan coefficients
of the conformal group. By known harmonic analysis \cite{key-10},
the coefficients are exactly three-point functions (in some arbitrary
normalization), namely, for a representation with a primary of spin
$J$ (for $d\geq4$, we may have to use to quantum numbers $\left(\ell,\overline{\ell}\right)$
for the spin representations, but we soon restrict our attention to
$d=3$ in which a single quantum number is sufficient) and dimension
$\Delta$ they are 
\begin{equation}
\left\langle \psi\left(x_{1}\right)\overline{\psi}\left(x_{2}\right)O^{\Delta,J}\left(x_{3}\right)\right\rangle .
\end{equation}

These have a few complexities that were absent in the scalar case:
\begin{enumerate}
\item In the scalar case we only had 3-pt. functions in which the scalars
coupled to symmetric traceless tensors (in $d=4$ these are $\left(\ell,\ell\right)$).
Fermions in $d\geq4$ can couple also to other operators, for example
in $d=4$ they can couple to $\left(\ell,\ell+2\right),\left(\ell+2,\ell\right)$.
\item In the scalar case we had two options for the scalar representations
- real or complex. Fermions can come in a few different representations,
such as Dirac, Majorana, (real or complex) Weyl. The different representations
(in addition to Dirac fermions) that are allowed also depend on the
spacetime dimension
\item Unlike in the scalar case, we now have four different tensor structures
(this is true for any dimension) that can appear in the 3-pt. function.
The 3-pt. function is given by 
\begin{equation}
\left\langle \psi\left(x_{1}\right)\overline{\psi}\left(x_{2}\right)O^{\Delta,J}\left(x_{3}\right)\right\rangle =\lambda_{\psi\overline{\psi}O}^{m}\left\langle \psi\left(x_{1}\right)\overline{\psi}\left(x_{2}\right)O^{\Delta,J}\left(x_{3}\right)\right\rangle ^{m}
\end{equation}
where $m=1,2,3,4$ runs over the different structures, which in turn
depend on the fermion representation.
\end{enumerate}
To deal with these complications, we will specify the dimension $d$
and the fermion representation. The simplest case would be the $d=3$
Majorana theory $\left(\overline{\psi}=\psi\right)$. These are in
fact expected to be continuously connected to the $d=3$ $O\left(N\right)$
critical scalar theory, and to have the same $O\left(N\right)$-invariant
operator spectrum (for free fermions at large $N$, after imposing
their equation of motion). The tensor structures in this case are
\cite{key-12} (Here and from now on we make use of the embedding
space formalism, and use for short $\left\langle \psi_{1}^{\Delta_{\psi}}\psi_{2}^{\Delta_{\psi}}O^{\Delta,J}\right\rangle =\left\langle \psi\left(X_{1},S_{1}\right)\psi\left(X_{2},S_{2}\right)O\left(X_{3},S_{3}\right)\right\rangle $.
Details can be found in appendix A)

\begin{align}
\left\langle \psi_{1}^{\Delta_{\psi}}\psi_{2}^{\Delta_{\psi}}O^{\Delta,J}\right\rangle ^{1} & =\frac{\left\langle S_{1}S_{2}\right\rangle \left\langle S_{3}X_{1}X_{2}S_{3}\right\rangle ^{J}}{X_{12}^{\left(2\Delta_{\psi}-\Delta+J+1\right)/2}X_{23}^{\left(\Delta+J\right)/2}X_{31}^{\left(\Delta+J\right)/2}},\\
\left\langle \psi_{1}^{\Delta_{\psi}}\psi_{2}^{\Delta_{\psi}}O^{\Delta,J}\right\rangle ^{2} & =\frac{\left\langle S_{2}S_{3}\right\rangle \left\langle S_{1}S_{3}\right\rangle \left\langle S_{3}X_{1}X_{2}S_{3}\right\rangle ^{J-1}}{X_{12}^{\left(2\Delta_{\psi}-\Delta+J-1\right)/2}X_{23}^{\left(\Delta+J\right)/2}X_{31}^{\left(\Delta+J\right)/2}},\\
\left\langle \psi_{1}^{\Delta_{\psi}}\psi_{2}^{\Delta_{\psi}}O^{\Delta,J}\right\rangle ^{3} & =\frac{\left\langle S_{3}X_{1}X_{2}S_{3}\right\rangle ^{J-1}\left\langle S_{1}S_{3}\right\rangle \left\langle S_{2}X_{1}S_{3}\right\rangle }{X_{12}^{\left(2\Delta_{\psi}-\Delta+J\right)/2}X_{23}^{\left(\Delta+J-1\right)/2}X_{31}^{\left(\Delta+J+1\right)/2}}+\frac{\left\langle S_{3}X_{1}X_{2}S_{3}\right\rangle ^{J-1}\left\langle S_{2}S_{3}\right\rangle \left\langle S_{1}X_{2}S_{3}\right\rangle }{X_{12}^{\left(2\Delta_{\psi}-\Delta+J\right)/2}X_{23}^{\left(\Delta+J+1\right)/2}X_{31}^{\left(\Delta+J-1\right)/2}},\\
\left\langle \psi_{1}^{\Delta_{\psi}}\psi_{2}^{\Delta_{\psi}}O^{\Delta,J}\right\rangle ^{4} & =\frac{\left\langle S_{3}X_{1}X_{2}S_{3}\right\rangle ^{J-1}\left\langle S_{1}S_{3}\right\rangle \left\langle S_{2}X_{1}S_{3}\right\rangle }{X_{12}^{\left(2\Delta_{\psi}-\Delta+J\right)/2}X_{23}^{\left(\Delta+J-1\right)/2}X_{31}^{\left(\Delta+J+1\right)/2}}-\frac{\left\langle S_{3}X_{1}X_{2}S_{3}\right\rangle ^{J-1}\left\langle S_{2}S_{3}\right\rangle \left\langle S_{1}X_{2}S_{3}\right\rangle }{X_{12}^{\left(2\Delta_{\psi}-\Delta+J\right)/2}X_{23}^{\left(\Delta+J+1\right)/2}X_{31}^{\left(\Delta+J-1\right)/2}}.
\end{align}
Where the first two structures are parity-even and the latter two
are parity-odd. Note that not all of these may appear independently
in correlation functions. First, for $J=0$ only the first and third
exist (the existence of the third may be seen by use of the Fierz
identities). Moreover, for operators with even $J$ we have $\lambda_{\psi\psi O^{-}}^{4}=0$
and for operators with odd $J$ we have $\lambda_{\psi\psi O^{+}}^{1}=\lambda_{\psi\psi O^{+}}^{2}=\lambda_{\psi\psi O^{-}}^{3}=0$
(here the $\pm$ in $O^{\pm}$ denotes parity). Also, since $\left\langle \psi_{1}^{\Delta_{\psi}}\psi_{2}^{\Delta_{\psi}}O^{\Delta,J}\right\rangle ^{*}=-\left\langle \psi_{1}^{\Delta_{\psi}}\psi_{2}^{\Delta_{\psi}}O^{\Delta,J}\right\rangle $,
all $\lambda_{\psi\psi O}^{m}$ must be pure imaginary. For a free
fermion ($\Delta_{\psi}=\frac{d-1}{2}=1$), the three point functions
must satisfy the EOM which will relate some of the coefficients
\begin{equation}
0=\cancel{\d}_{1}\left\langle \psi^{\Delta_{\psi}=1}\left(x_{1}\right)\psi^{\Delta_{\psi}=1}\left(x_{2}\right)O^{\Delta,J}\left(x_{3}\right)\right\rangle =\lambda_{\psi\psi O}^{m}\cancel{\d}_{1}\left\langle \psi^{\Delta_{\psi}=1}\left(x_{1}\right)\psi^{\Delta_{\psi}=1}\left(x_{2}\right)O^{\Delta,J}\left(x_{3}\right)\right\rangle ^{m},
\end{equation}
and this relation is useful for showing that on the gravity side one
gets on-shell a single tower of higher-spin degrees of freedom as
expected for Vasiliev gravity. We can use the embedding space notations
we introduced above to calculate the physical space expressions for
the tensor structures. For example 
\begin{align}
\left\langle S_{1}S_{2}\right\rangle  & =S_{1I}\Omega^{IJ}S_{2J}=s_{2\alpha}x_{1}^{\mu}\gamma_{\mu\beta}^{\alpha}s_{1}^{\beta}-s_{1\alpha}x_{2}^{\mu}\gamma_{\mu\beta}^{\alpha}s_{2}^{\beta},\\
\left\langle S_{1}X_{2}S_{3}\right\rangle  & =4x_{12}^{\mu}x_{32}^{\nu}s_{1}\left(\frac{1}{4}\left[\gamma_{\mu},\gamma_{\nu}\right]\right)s_{3},\\
\left\langle S_{3}X_{1}X_{2}S_{3}\right\rangle  & =x_{13}^{2}x_{23}^{2}\left(\frac{x_{23}^{\mu}}{x_{23}^{2}}-\frac{x_{13}^{\mu}}{x_{13}^{2}}\right)s_{3}\gamma_{\mu}s_{3},\\
\left\langle \psi^{\alpha}\left(x_{1}\right)\psi^{\beta}\left(x_{2}\right)O_{J}^{\mu_{1}\dots\mu_{J}}\left(x_{3}\right)\right\rangle ^{1} & =\frac{\left(-1\right)^{J}}{2^{J}\left(2J\right)!}\gamma_{\alpha_{1}\alpha_{2}}^{\mu_{1}}\cdots\gamma_{\alpha_{2J-1,\alpha_{2J}}}^{\mu_{J}}\frac{\d^{2J}}{\d s_{\alpha_{1}}\cdots\d s_{\alpha_{2J}}}\cross\nonumber \\
 & \qquad\cross\left(\frac{x_{13}^{2}x_{23}^{2}}{x_{12}^{2}}\left(\frac{x_{23}^{\mu}}{x_{23}^{2}}-\frac{x_{13}^{\mu}}{x_{13}^{2}}\right)s\gamma_{\mu}s\right)^{J}\frac{x_{12}^{\mu}\gamma_{\mu}^{\alpha\beta}}{x_{12}^{2\Delta_{\psi}-\Delta-J+1}x_{23}^{\Delta+J}x_{31}^{\Delta+J}},
\end{align}
etc. We define the pairing of $n$-pt. functions as 

\begin{equation}
\left(\MO\left(x_{1},\dots,x_{n}\right),\MO^{\tg}\left(x_{1},\dots,x_{n}\right)\right)=\intop\frac{d^{d}x_{1}\cdots d^{d}x_{n}}{\text{vol}\left(SO\left(d+1,1\right)\right)}\MO\left(x_{1},\dots,x_{n}\right)\MO^{\tg}\left(x_{1},\dots,x_{n}\right).
\end{equation}

The pairing $\left(\left\langle \psi_{1}^{\Delta_{1}}\psi_{2}^{\Delta_{2}}O^{\Delta,J}\right\rangle ^{m},\left\langle \psi_{1}^{\tilde{\Delta}_{1}}\psi_{2}^{\tilde{\Delta}_{2}}O^{\tilde{\Delta},J}\right\rangle ^{n}\right)$
of 3-pt. functions is calculated in appendix B.1. We can use it to
define the bubble coefficients 
\begin{equation}
\MB_{\Delta,J}^{mn}=\frac{\left(\left\langle \psi_{1}\psi_{2}O^{\Delta,J}\right\rangle ^{m},\left\langle \tilde{\psi}_{1}\tilde{\psi}_{2}O^{\tilde{\Delta},J}\right\rangle ^{n}\right)}{\mu\left(\Delta,J\right)}
\end{equation}
where $\mu\left(\Delta,J\right)$ is the Plancherel measure for the
representation $\Delta,J$ of the Euclidean conformal group $SO\left(4,1\right)$.
The Plancherel measure for a group $G$ is a measure on the space
of unitary irreducible representations of $G.$ For compact groups
and for a unitary irreducible rep. $\pi$, it is given by 
\begin{equation}
\mu\left(\pi\right)=\frac{\text{dim}\pi}{\text{Vol}G}
\end{equation}
where $\text{Vol}G$ is defined using the Haar measure on $G$. For
non-compact groups, both $\text{dim}\pi$ and $\text{Vol}G$ can be
infinite, and one can think of the Plancherel measure as a regularized
quotient. In our case one can calculate and find 
\begin{equation}
\mu\left(\Delta,J\right)=\frac{\left(2\Delta-3\right)\left(J+\Delta-1\right)\left(J+\tilde{\Delta}-1\right)\Gamma\left(2J+2\right)\cot\left(\pi\left(\Delta+J\right)\right)}{128\pi^{5}}.
\end{equation}
 We can write an expression for the orthogonality of the 3-pt. functions
\begin{align}
 & \intop d^{d}x_{1}d^{d}x_{2}\frac{\d^{2}}{\d s_{1\alpha}\d s_{1}^{\alpha}}\frac{\d^{2}}{\d s_{2\beta}\d s_{2}^{\beta}}\left\langle \psi\left(x_{1},s_{1}\right)\psi\left(x_{2},s_{2}\right)O^{\Delta,J}\left(y,s\right)\right\rangle ^{m}\left\langle \tilde{\psi}\left(x_{1},s_{1}\right)\tilde{\psi}\left(x_{2},s_{2}\right)O^{\tilde{\Delta^{\tg}},J^{\tg}}\left(y^{\tg},s^{\tg}\right)\right\rangle ^{n}\nonumber \\
 & \qquad=2\pi i\MB_{\Delta,J}^{mk}\delta_{JJ^{\tg}}\left(\delta_{k}^{n}\delta\left(\Delta-\Delta^{\tg}\right)\delta^{\left(d\right)}\left(y-y^{\tg}\right)\left(s\cdot s^{\tg}\right)^{J}+\left(S_{k}^{n}\left(\psi\psi\left[\MO_{\Delta,J}\right]\right)\right)^{-1}\delta\left(\Delta-\tilde{\Delta^{\tg}}\right)\left\langle O^{\Delta,J}\left(y,s\right)O^{\Delta,J}\left(y^{\tg},s^{\tg}\right)\right\rangle \right)\label{eq:ortho}
\end{align}

where $S_{k}^{n}\left(\psi\psi\left[\MO_{\Delta,J}\right]\right)$
are the shadow coefficients whose explicit expressions are given in
appendix B.2. Our bilocal fields can then be expressed as the expansion
\begin{equation}
\rho_{\alpha\beta}\left(x_{1},x_{2}\right)=\sum_{m=1}^{4}\sum_{J=0}^{\infty}\intop_{C_{J}}\frac{d\Delta}{2\pi i}\intop d^{d}yc_{\Delta,J}^{m}\left(y,\d_{s}\right)\left\langle \psi_{\alpha}\psi_{\beta}O^{\Delta,J}\left(y,s\right)\right\rangle ^{m}\label{eq:rhoab}
\end{equation}
which is invertible by 
\begin{equation}
c_{\Delta,J}^{m}\left(y,s\right)=\left(\MB_{\Delta,J}^{mn}\right)^{-1}\intop d^{d}x_{1}d^{d}x_{2}\rho^{\alpha\beta}\left(x_{1},x_{2}\right)\left\langle \tilde{\psi}_{\alpha}\tilde{\psi}_{\beta}O^{\tilde{\Delta},J}\left(y,s\right)\right\rangle ^{n}.\label{eq:cdj}
\end{equation}

Note that the coefficients $c_{\Delta,J}^{m}$ have a spinor index
which is encoded above using the auxiliary variable $s$. The $\d_{s}$
argument contracts this index with the spinor indices of the operator
$O^{\Delta,J}$. We will suppress these indices in the following analysis
to avoid cluttering of the equations. Completeness is given by 
\begin{equation}
\delta\left(x_{13}\right)\delta\left(x_{24}\right)\Omega_{\alpha\gamma}\Omega_{\beta\delta}=\sum_{m,n1}^{4}\sum_{J=0}^{\infty}\intop_{C_{J}}\frac{d\Delta}{2\pi i}\intop d^{d}y\left(\MB_{\Delta,J}^{mn}\right)^{-1}\left\langle \psi_{\alpha}\psi_{\beta}O^{\Delta,J}\right\rangle ^{m}\left\langle \tilde{\psi}_{\gamma}\tilde{\psi}_{\delta}O^{\tilde{\Delta},J}\right\rangle ^{n}.
\end{equation}

The second term in the orthogonality theorem tells us that there is
a relation between the basis elements for $\Delta$ and $\tilde{\Delta}$.
They are not independent of each other, but instead are related by
the shadow transform 
\begin{equation}
\left\langle \psi_{\alpha}\left(x_{1}\right)\psi_{\beta}\left(x_{2}\right)O^{\Delta,J}\left(y\right)\right\rangle ^{m}=\left(S_{n}^{m}\left(\psi\psi\left[\MO_{\Delta,J}\right]\right)\right)^{-1}\intop d^{d}y^{\tg}\left\langle O^{\Delta,J}\left(y\right)O^{\Delta,J}\left(y^{\tg}\right)\right\rangle \left\langle \psi_{\alpha}\left(x_{1}\right)\psi_{\beta}\left(x_{2}\right)O^{\tilde{\Delta},J}\left(y^{\tg}\right)\right\rangle ^{n}.
\end{equation}

Which gives us a relation between the coefficients $c_{\Delta,J}^{m}$
and their shadow counterparts 
\begin{equation}
c_{\tilde{\Delta},J}^{m}\left(y\right)=\left(S_{n}^{m}\left(\psi\psi\left[\MO_{\Delta,J}\right]\right)\right)^{-1}\intop d^{d}y^{\tg}\left\langle O^{\Delta,J}\left(y\right)O^{\Delta,J}\left(y^{\tg}\right)\right\rangle c_{\Delta,J}^{n}\left(y^{\tg}\right).\label{eq:shadow}
\end{equation}

A subtlety of this construction is that the contour $C_{J}$ we used
in the expansion of $\rho_{\alpha\beta}$ (\ref{eq:rhoab}) and the
completeness relation, was shown in the literature to hold on the
principal series $\Delta=d/2+is$ with $s\in\BR$. However, we would
like to use it for real values of $\Delta_{\psi}$ as those are the
conformal weights of our physical operators. We can do this by analytically
continuing our expressions to real values of $\Delta$, but we need
to carefully check for which values of $\Delta,\Delta_{\psi}$ the
integral in (\ref{eq:cdj}) still converges. The convergence conditions
are:
\begin{enumerate}
\item In the limit of large $x_{1}$ (or equivalently large $x_{2}$) we
use $\rho_{\alpha\beta}\sim\left|x_{1}\right|^{-2\Delta_{\psi}}$
and $\left\langle \tilde{\psi}_{\alpha}\tilde{\psi}_{\beta}O^{\tilde{\Delta},J}\right\rangle \sim\left|x_{1}\right|^{-2\tilde{\Delta}_{\psi}}$
and get powers of $\left|x_{1}\right|$ that are equal to $d-2\text{Re}\left(\Delta_{\psi}\right)-2\text{Re}\left(\tilde{\Delta}_{\psi}\right)=-d$,
which is convergent for any value of $\Delta_{\psi}$.
\item In the limit $x_{1}\sim x_{2}$ (assuming a finite limit of $\rho_{\alpha\beta}\left(x_{1},x_{2}\right)$)
we get powers of $\left|x_{1}-x_{2}\right|$ equal to $d+\text{Re}\left(\tilde{\Delta}\right)-2\text{Re}\left(\tilde{\Delta}_{\psi}\right)=2\text{Re}\left(\Delta_{\psi}\right)-\text{Re}\left(\Delta\right)$
which for $\Delta=d/2+is$ gives the convergence condition $\Delta_{\psi}>d/4$.
\end{enumerate}
It is clear that for the free fermions where $\Delta_{\psi}=\frac{d-1}{2}$
there is no need to worry about convergence in $d>2$. In the scalar
case, one had to define a double shadow transform of the bilocal field
$\eta$ to ensure convergence. In the fermion case there is no need
to worry about convergence as we have seen, but we will keep these
definitions to get a nicer expression for our quadratic action, as
we will see in the following subsection. We define a new function
$\tilde{\rho}_{\alpha\beta}$ which is the double shadow transform
of $\rho_{\alpha\beta}$ by 
\begin{equation}
\rho_{\alpha\beta}\left(x_{1},x_{2}\right)=\intop d^{d}xd^{d}x^{\tg}\left\langle \psi_{\alpha}^{\Delta_{\psi}}\left(x_{1}\right)\psi_{\gamma}^{\Delta_{\psi}}\left(x\right)\right\rangle \left\langle \psi_{\beta}^{\Delta_{\psi}}\left(x_{2}\right)\psi_{\delta}^{\Delta_{\psi}}\left(x^{\tg}\right)\right\rangle \tilde{\rho}^{\gamma\delta}\left(x,x^{\tg}\right).\label{eq:rhotilde1}
\end{equation}
Now, $\tilde{\rho}_{\alpha\beta}$ transforms under the tensor product
of two $\tilde{\Delta}_{\psi}$ representations, and obeys the same
boundary conditions as $\rho_{\alpha\beta}$, except replacing $\Delta_{\psi}$
with $\tilde{\Delta}_{\psi}$. We can now expand 
\begin{equation}
\tilde{\rho}_{\alpha\beta}\left(x_{1},x_{2}\right)=\sum_{m=1}^{4}\sum_{J=0}^{\infty}\intop_{C_{J}}\frac{d\Delta}{2\pi i}\intop d^{d}y\tilde{c}_{\Delta,J}^{m}\left(y\right)\left\langle \tilde{\psi}_{\alpha}\tilde{\psi}_{\beta}O^{\Delta,J}\left(y\right)\right\rangle ^{m}\label{eq:rhotilde2}
\end{equation}
 with the coefficients 
\begin{equation}
\tilde{c}_{\Delta,J}^{m}\left(y\right)=\left(\MB_{\Delta,J}^{mn}\right)^{-1}\intop d^{d}x_{1}d^{d}x_{2}\tilde{\rho}^{\alpha\beta}\left(x_{1},x_{2}\right)\left\langle \psi_{\alpha}\psi_{\beta}O^{\tilde{\Delta},J}\left(y\right)\right\rangle ^{n}.
\end{equation}

Where the integral is now convergent. We can now plug in (\ref{eq:rhotilde2})
into (\ref{eq:rhotilde1}) to get 
\begin{align}
\rho_{\alpha\beta}\left(x_{1},x_{2}\right) & =\sum_{m=1}^{4}\sum_{J=0}^{\infty}\intop_{C_{J}}\frac{d\Delta}{2\pi i}\intop d^{d}y\intop d^{d}xd^{d}x^{\tg}\left\langle \psi_{\alpha}^{\Delta_{\psi}}\left(x_{1}\right)\psi_{\gamma}^{\Delta_{\psi}}\left(x\right)\right\rangle \left\langle \psi_{\beta}^{\Delta_{\psi}}\left(x_{2}\right)\psi_{\delta}^{\Delta_{\psi}}\left(x^{\tg}\right)\right\rangle \cross\nonumber \\
 & \quad\cross\tilde{c}_{\Delta,J}^{m}\left(y\right)\left\langle \tilde{\psi}^{\gamma}\left(x\right)\tilde{\psi}^{\delta}\left(x^{\tg}\right)O^{\Delta,J}\left(y\right)\right\rangle ^{m}\\
 & =\sum_{m=1}^{4}\sum_{J=0}^{\infty}\intop_{C_{J}}\frac{d\Delta}{2\pi i}\intop d^{d}y\tilde{c}_{\Delta,J}^{m}\left(y\right)S_{n}^{m}\left(\left[\tilde{\psi}\right]\tilde{\psi}\MO^{\Delta,J}\right)S_{k}^{n}\left(\psi\left[\tilde{\psi}\right]\MO^{\Delta,J}\right)\left\langle \psi_{\alpha}\left(x_{1}\right)\psi_{\beta}\left(x_{2}\right)O^{\Delta,J}\left(y\right)\right\rangle ^{k},
\end{align}
where we used twice the shadow transform (with the short hand notation
$\psi=\psi^{\Delta_{\psi}},\ \tilde{\psi}=\psi^{\tilde{\Delta}_{\psi}}$)
\begin{equation}
\intop d^{d}x\left\langle \psi_{\alpha}^{\Delta_{1}}\left(x_{1}\right)\psi_{\gamma}^{\Delta_{1}}\left(x\right)\right\rangle \left\langle \psi^{\tilde{\Delta}_{1}\gamma}\left(x\right)\psi^{\Delta_{2}\delta}\left(x^{\tg}\right)O^{\Delta,J}\left(y\right)\right\rangle ^{m}=S_{n}^{m}\left(\left[\psi^{\tilde{\Delta}_{1}}\right]\psi^{\Delta_{2}}\MO^{\Delta,J}\right)\left\langle \psi^{\Delta_{1}\gamma}\left(x\right)\psi^{\Delta_{2}\delta}\left(x^{\tg}\right)O^{\Delta,J}\left(y\right)\right\rangle ^{n}.\label{eq:shadowtrans}
\end{equation}

Note that the shadow coefficients we need here are different than
the ones we defined earlier. However, they can be obtained from the
literature \cite{key-11} and appear explicitly in appendix B.3. Note
that we get a block anti-diagonal form for the shadow coefficients
due to parity. We then get that the matrix 
\begin{equation}
A_{k;\Delta,J}^{m}=S_{n}^{m}\left(\left[\tilde{\psi}\right]\tilde{\psi}\MO^{\Delta,J}\right)S_{k}^{n}\left(\psi\left[\tilde{\psi}\right]\MO^{\Delta,J}\right)
\end{equation}
is block diagonal, with the blocks 
\begin{equation}
A_{k;\Delta,J}^{m}=\begin{cases}
\frac{\pi^{3}\Gamma^{2}\left(2-\Delta_{\psi}\right)\Gamma\left(\Delta_{\psi}-\frac{\Delta-J+1}{2}\right)\Gamma\left(\Delta_{\psi}-\frac{\tilde{\Delta}-J+1}{2}\right)}{\Gamma^{2}\left(\Delta_{\psi}+\frac{1}{2}\right)\Gamma\left(\tilde{\Delta}_{\psi}-\frac{\Delta-J-1}{2}\right)\Gamma\left(\tilde{\Delta}_{\psi}-\frac{\tilde{\Delta}-J-1}{2}\right)}\cross\\
\quad\cross\left(\begin{array}{cc}
-\left(\tilde{\Delta}_{\psi}-\frac{\Delta-J+2}{2}\right)\left(\Delta_{\psi}-\frac{\Delta-J+1}{2}\right) & J(\Delta-1)\\
\left(\tilde{\Delta}_{\psi}-\frac{\Delta-J+1}{2}\right)\left(\Delta_{\psi}-\frac{\Delta-J+1}{2}\right) & \left(\tilde{\Delta}_{\psi}-\frac{\Delta-J+1}{2}\right)\left(\Delta_{\psi}-\frac{\Delta-J+2}{2}\right)
\end{array}\right) & m,k=1,2\\
\frac{\pi^{3}\Gamma^{2}\left(2-\Delta_{\psi}\right)\Gamma\left(\Delta_{\psi}-\frac{\Delta-J}{2}\right)\Gamma\left(\Delta_{\psi}-\frac{\tilde{\Delta}-J}{2}\right)}{\Gamma^{2}\left(\Delta_{\psi}+\frac{1}{2}\right)\Gamma\left(\tilde{\Delta}_{\psi}-\frac{\Delta-J}{2}\right)\Gamma\left(\tilde{\Delta}_{\psi}-\frac{\tilde{\Delta}-J}{2}\right)}\begin{pmatrix}1 & 0\\
0 & -1
\end{pmatrix} & m,k=3,4
\end{cases}.\label{eq:Amk}
\end{equation}

We can now analytically continue our expression for $\rho_{\alpha\beta}$,
with the only problem coming from the poles of $\left(A_{k;\Delta,J}^{m}\right)^{-1}$.
For odd parity operators $O^{\Delta,J}$ this function has the pole
structure we expect from the dimensions of the physical primary operators
of the generalized free field theory
\begin{equation}
\Delta^{\left(n,J\right)}=2\Delta_{\psi}+2n+J
\end{equation}
and their appropriate $\tilde{\Delta}^{\left(n,J\right)}$. For even
parity operators the pole structure is a bit more complicated. The
idea is that we can now deform the contour $C_{J}$ to a new one $\gamma_{J}$
which includes the poles $\Delta^{\left(n,J\right)}$ that are to
the right of the principal series ($\text{Re}\left(\Delta\right)=d/2$)
and excludes the poles $\tilde{\Delta}^{\left(n,J\right)}$ to its
left (which appear at high enough $n,J$). We then get a convergent
definition for 
\begin{equation}
\rho_{\alpha\beta}\left(x_{1},x_{2}\right)=\sum_{m=1}^{4}\sum_{J=0}^{\infty}\intop_{\gamma_{J}}\frac{d\Delta}{2\pi i}\intop d^{d}yc_{\Delta,J}^{m}\left(y\right)\left\langle \psi_{\alpha}\psi_{\beta}O^{\Delta,J}\left(y\right)\right\rangle ^{m}
\end{equation}
where we define 
\begin{equation}
c_{\Delta,J}^{m}=A_{k;\Delta,J}^{m}\tilde{c}_{\Delta,J}^{k}.
\end{equation}

We now turn our attention to the free fermion case, for which $\Delta_{\psi}=\frac{d-1}{2}$,
and in our case $d=3$ and $\Delta_{\psi}=1,\tilde{\Delta}_{\psi}=2$.
Here we have 
\begin{equation}
A_{k;\Delta,J}^{m}=\begin{cases}
\frac{\left(4\pi\right)^{2}}{\left(1-\Delta+J\right)\left(1-\tilde{\Delta}+J\right)\left(\Delta+J\right)\left(\tilde{\Delta}+J\right)}\left(\begin{array}{cc}
-\left(2-\Delta+J\right)\left(1-\Delta+J\right) & 4J(\Delta-1)\\
\left(\tilde{\Delta}+J\right)\left(1-\Delta+J\right) & \left(\tilde{\Delta}+J\right)\left(-\Delta+J\right)
\end{array}\right) & m,k=1,2\\
\frac{\left(4\pi\right)^{2}}{\left(2-\Delta+J\right)\left(2-\tilde{\Delta}+J\right)}\begin{pmatrix}1 & 0\\
0 & -1
\end{pmatrix} & m,k=3,4
\end{cases}.
\end{equation}

And taking the inverse we get 
\begin{equation}
\left(A_{\Delta,J}^{-1}\right)_{\,k}^{m}=\frac{1}{\left(4\pi\right)^{2}}\begin{cases}
\left(\begin{array}{cc}
-\left(3-\Delta+J\right)\left(-\Delta+J\right) & 4J\left(\Delta-1\right)\\
\left(3-\Delta+J\right)\left(1-\Delta+J\right) & \ \left(2-\Delta+J\right)\left(1-\Delta+J\right)
\end{array}\right) & m,k=1,2\\
\left(\begin{array}{cc}
\left(2-\Delta+J\right)\left(-1+\Delta+J\right) & 0\\
0 & -\left(2-\Delta+J\right)\left(-1+\Delta+J\right)
\end{array}\right) & m,k=3,4
\end{cases}.
\end{equation}

\subsection{The quadratic bilocal action}

We can now write our quadratic bilocal action (\ref{eq:quadaction})
in this formalism 
\begin{align}
S_{quad} & =-\frac{1}{2}\intop d^{d}x_{1}d^{d}x_{2}\rho_{\gamma\delta}\left(x_{1},x_{2}\right)\cancel{\d}_{1\beta}^{\gamma}\cancel{\d}_{2\alpha}^{\delta}\rho^{\alpha\beta}\left(x_{1},x_{2}\right)\nonumber \\
 & =-\frac{1}{2}\intop d^{d}x_{1}d^{d}x_{2}\tilde{\rho}_{\gamma\delta}\left(x_{1},x_{2}\right)x_{12}^{2}\cancel{\d}_{1\beta}^{\gamma}\cancel{\d}_{2\alpha}^{\delta}\rho^{\alpha\beta}\left(x_{1},x_{2}\right)\nonumber \\
 & =-\frac{1}{2}\sum_{m,m^{\tg}=1}^{4}\sum_{J,J^{\tg}=0}^{\infty}\intop_{\gamma_{J}}\frac{d\Delta}{2\pi i}\intop_{\gamma_{J^{\tg}}}\frac{d\Delta^{\tg}}{2\pi i}\intop d^{d}yd^{d}y^{\tg}\tilde{c}_{\Delta^{\tg},J^{\tg}}^{m^{\tg}}c_{\Delta,J}^{m}\cross\nonumber \\
 & \ \ \cross\intop d^{d}x_{1}d^{d}x_{2}\left\langle \tilde{\psi}^{\gamma}\tilde{\psi}^{\delta}O^{\Delta^{\tg},J^{\tg}}\right\rangle ^{m^{\tg}}x_{12}^{2}\cancel{\d}_{1\beta}^{\gamma}\cancel{\d}_{2\alpha}^{\delta}\left\langle \psi^{\alpha}\psi^{\beta}O^{\Delta,J}\right\rangle ^{m}.
\end{align}
One can note here the quadratic action operator 
\begin{equation}
S_{qa}=x_{12}^{2}\cancel{\d}_{1\beta}^{\gamma}\cancel{\d}_{2\alpha}^{\delta}.
\end{equation}

This operator is manifestly conformally invariant, as can be checked
explicitly by commuting it with the bilocal conformal generators.
Acting with $S_{qa}$ on the 3-pt. function basis with $\Delta_{\psi}=1$
was done using Wolfram Mathematica. The code is given in appendix
C. it gives the result 
\begin{align}
\left(S_{qa}\right)_{\gamma\delta}^{\alpha\beta}\left\langle \psi_{\alpha}\psi_{\beta}O^{\Delta,J}\right\rangle ^{m} & \equiv B_{k;\Delta,J}^{m}\left\langle \psi_{\gamma}\psi_{\delta}O^{\Delta,J}\right\rangle ^{k}\nonumber \\
 & =-\left(4\pi\right)^{2}\left(A_{\Delta,J}^{-1}\right)_{\,k}^{m}\left\langle \psi_{\gamma}\psi_{\delta}O^{\Delta,J}\right\rangle ^{k}.\label{eq:sqa1}
\end{align}
We see that the action matrix in the 3-pt. basis is exactly the same
as the matrix which relates between $c_{\Delta,J}^{m},\tilde{c}_{\Delta,J}^{k}$.
We can use this fact to derive the relation 
\begin{align}
\tilde{\rho}_{\alpha\beta}\left(x_{1},x_{2}\right) & =\sum_{m=1}^{4}\sum_{J=0}^{\infty}\intop_{\gamma_{J}}\frac{d\Delta}{2\pi i}\intop d^{d}y\tilde{c}_{\Delta,J}^{m}\left\langle \tilde{\psi}_{\alpha}\tilde{\psi}_{\beta}O^{\Delta,J}\right\rangle ^{m}\nonumber \\
 & =\sum_{m=1}^{4}\sum_{J=0}^{\infty}\intop_{\gamma_{J}}\frac{d\Delta}{2\pi i}\intop d^{d}yc_{\Delta,J}^{m}\left(A_{\Delta,J}^{-1}\right)^{mk}\left\langle \tilde{\psi}_{\alpha}\tilde{\psi}_{\beta}O^{\Delta,J}\right\rangle ^{k}\nonumber \\
 & =\frac{1}{x_{12}^{2}}\sum_{m=1}^{4}\sum_{J=0}^{\infty}\intop_{\gamma_{J}}\frac{d\Delta}{2\pi i}\intop d^{d}yc_{\Delta,J}^{m}\left(A_{\Delta,J}^{-1}\right)^{mk}\left\langle \psi_{\alpha}\psi_{\beta}O^{\Delta,J}\right\rangle ^{k}\nonumber \\
 & =\frac{-1}{\left(4\pi\right)^{2}}\cancel{\d}_{1\beta}^{\gamma}\cancel{\d}_{2\alpha}^{\delta}\rho_{\gamma\delta}\left(x_{1},x_{2}\right).
\end{align}

Note that this also follows directly from (\ref{eq:rhotilde1}). We
can write our quadratic action in terms of the coefficients 
\begin{align}
S_{quad}\left[c_{\Delta,J}^{m}\right] & =\frac{1}{2\left(4\pi\right)^{2}}\intop d^{d}x_{1}d^{d}x_{2}\rho_{\alpha\beta}\left(x_{1},x_{2}\right)\tilde{\rho}^{\alpha\beta}\left(x_{1},x_{2}\right)\nonumber \\
 & =\frac{1}{2\left(4\pi\right)^{2}}\sum_{J,J^{\tg}=0}^{\infty}\intop_{\gamma_{J}}\frac{d\Delta}{2\pi i}\intop_{\gamma_{J^{\tg}}}\frac{d\Delta^{\tg}}{2\pi i}\intop d^{d}yd^{d}y^{\tg}c_{\Delta,J}^{m}\left(A_{\tilde{\Delta}^{\tg},J}^{-1}\right)_{m^{\tg}k}c_{\tilde{\Delta}^{\tg},J^{\tg}}^{k}\cross\nonumber \\
 & \ \cross\intop d^{d}x_{1}d^{d}x_{2}\left\langle \psi_{\alpha}\psi_{\beta}O^{\Delta,J}\right\rangle ^{m}\left\langle \tilde{\psi}^{\alpha}\tilde{\psi}^{\beta}O^{\tilde{\Delta}^{\tg},J^{\tg}}\right\rangle ^{m^{\tg}}\\
 & =\frac{1}{\left(4\pi\right)^{2}}\sum_{J=0}^{\infty}\intop_{\gamma_{J}}\frac{d\Delta}{2\pi i}\intop d^{d}yc_{\Delta,J}^{m}\MB_{\Delta,J}^{mm^{\tg}}\left(A_{\Delta,J}^{-1}\right)_{m^{\tg}k}c_{\tilde{\Delta},J}^{k}.\label{eq:squad}
\end{align}

Where in the second equality we used the relation $c_{\Delta,J}^{m}=A_{k;\Delta,J}^{m}\tilde{c}_{\Delta,J}^{k}$,
and in the last equality we used the orthogonality relation (\ref{eq:ortho}).

\subsection{Bilocal theory 2-point function}

We can now calculate the bilocal theory 2-pt. function, and see that
it indeed reproduces the 4-pt. function of the free theory, as must
be the case since 
\begin{align}
\left\langle \rho_{\alpha\beta}\left(x_{1},x_{2}\right)\rho_{\gamma\delta}\left(x_{3},x_{4}\right)\right\rangle  & =\frac{1}{N}\left\langle \left(\psi_{\alpha}^{i}\left(x_{1}\right)\psi_{\beta}^{i}\left(x_{2}\right)-\left\langle \psi_{\alpha}^{i}\left(x_{1}\right)\psi_{\beta}^{i}\left(x_{2}\right)\right\rangle \right)\left(\psi_{\gamma}^{j}\left(x_{3}\right)\psi_{\delta}^{j}\left(x_{4}\right)-\left\langle \psi_{\gamma}^{j}\left(x_{3}\right)\psi_{\delta}^{j}\left(x_{4}\right)\right\rangle \right)\right\rangle \nonumber \\
 & =\frac{1}{N}\left(\left\langle \psi_{\alpha}^{i}\left(x_{1}\right)\psi_{\beta}^{i}\left(x_{2}\right)\psi_{\gamma}^{j}\left(x_{3}\right)\psi_{\delta}^{j}\left(x_{4}\right)\right\rangle -\left\langle \psi_{\alpha}^{i}\left(x_{1}\right)\psi_{\beta}^{i}\left(x_{2}\right)\right\rangle \left\langle \psi_{\gamma}^{j}\left(x_{3}\right)\psi_{\delta}^{j}\left(x_{4}\right)\right\rangle \right).
\end{align}

Using only the quadratic term in the action, we have 
\begin{align}
\left\langle \rho_{\alpha\beta}\left(x_{1},x_{2}\right)\rho_{\gamma\delta}\left(x_{3},x_{4}\right)\right\rangle _{\text{tree}} & =\sum_{J,J^{\tg}=0}^{\infty}\intop_{\gamma_{J}}\frac{d\Delta}{2\pi i}\intop_{\gamma_{J^{\tg}}}\frac{d\Delta^{\tg}}{2\pi i}\intop d^{d}yd^{d}y^{\tg}\left\langle \psi_{\alpha}\psi_{\beta}O^{\Delta,J}\right\rangle ^{m}\left\langle \psi^{\alpha}\psi^{\beta}O^{\tilde{\Delta}^{\tg},J^{\tg}}\right\rangle ^{m^{\tg}}\cross\nonumber \\
 & \ \cross\intop\left(dc_{\Delta,J}^{m}\right)c_{\Delta,J}^{m}c_{\tilde{\Delta}^{\tg},J^{\tg}}^{m^{\tg}}e^{-S_{quad}\left[c_{\Delta,J}^{m}\right]}.
\end{align}

This is a Gaussian path integral. It is useful to define the conformal
partial wave \cite{key-2} 
\begin{equation}
\Psi_{\Delta,J,\alpha\beta\gamma\delta}^{\Delta_{\psi}\left(mk\right)}\left(x_{1},\dots,x_{4}\right)=\intop d^{d}y\left\langle \psi_{\alpha}\psi_{\beta}O^{\Delta,J}\left(y\right)\right\rangle ^{m}\left\langle \psi_{\gamma}\psi_{\delta}O^{\tilde{\Delta},J}\left(y\right)\right\rangle ^{k}.
\end{equation}
So we get 
\begin{align}
\left\langle \rho_{\alpha\beta}\left(x_{1},x_{2}\right)\rho_{\gamma\delta}\left(x_{3},x_{4}\right)\right\rangle _{\text{tree}} & =\sum_{J=0}^{\infty}\intop_{\gamma_{J}}\frac{d\Delta}{2\pi i}\MB_{\Delta,J}^{mm^{\tg}}\left(A_{\Delta,J}^{-1}\right)_{m^{\tg}k}\Psi_{\Delta,J,\alpha\beta\gamma\delta}^{\Delta_{\psi}\left(mk\right)}\left(x_{1},\dots,x_{4}\right).
\end{align}

Recalling our derivation (\ref{eq:Amk}), one can note that the coefficients
$\MB_{\Delta,J}^{mm^{\tg}}\left(A_{\Delta,J}^{-1}\right)_{m^{\tg}k}$
are exactly what one would get from bootstrapping the local fermionic
MFT, without going through the bilocal theory. This calculation was
done in detail by \cite{key-12,key-2} and it indeed gives the fermionic
free field 4-pt. function as expected 
\begin{equation}
\left\langle \rho_{\alpha\beta}\left(x_{1},x_{2}\right)\rho_{\gamma\delta}\left(x_{3},x_{4}\right)\right\rangle =\left(\frac{\Gamma\left(\frac{d}{2}\right)}{2\pi^{d/2}}\right)^{2}\left(\frac{\gamma_{\alpha\gamma}^{\mu}x_{13\mu}}{\left|x_{13}\right|^{d}}\frac{\gamma_{\beta\delta}^{\nu}x_{24\nu}}{\left|x_{24}\right|^{d}}-\frac{\gamma_{\alpha\delta}^{\mu}x_{14\mu}}{\left|x_{14}\right|^{d}}\frac{\gamma_{\beta\gamma}^{\nu}x_{23\nu}}{\left|x_{23}\right|^{d}}\right).
\end{equation}

\subsection{Mapping to AdS}

We now attempt to map our bilocal degrees of freedom into AdS spacetime.
In the bosonic field theory this was achieved by mapping the eigenvalues
of the action $\lambda_{\Delta,J}$ to the masses of AdS fields $M_{\Delta,J}$.
In the bosonic free theory 
\begin{align}
\lambda_{\Delta,J} & =\left(\Delta+J\right)\left(d-\Delta+J\right)\left(d-2-\Delta+J\right)\left(\Delta+J-2\right)\nonumber \\
 & =\left(M_{\Delta,J}^{2}-M_{d+J,J}^{2}\right)\left(M_{\Delta,J}^{2}-M_{d+J-2,J}^{2}\right),\label{eq:lambdadj}
\end{align}
where 
\begin{equation}
M_{\Delta,J}^{2}=\Delta\left(\Delta-d\right)-J.
\end{equation}

The analogue of $\lambda_{\Delta,J}$ in the fermionic free theory
is $\left(\lambda_{\Delta,J}\right)_{\,k}^{m}=\left(4\pi\right)^{2}\left(A_{\Delta,J}^{-1}\right)_{\,k}^{m}$.
Diagonalizing it we get 
\begin{equation}
\left(\lambda_{\Delta,J}\right)_{\,k}^{m}=\begin{cases}
\left(\begin{array}{cc}
-\left(3-\Delta+J\right)\left(-\Delta+J\right) & 4J\left(\Delta-1\right)\\
\left(3-\Delta+J\right)\left(1-\Delta+J\right) & \ \left(2-\Delta+J\right)\left(1-\Delta+J\right)
\end{array}\right) & m,k=1,2\\
\left(\begin{array}{cc}
\left(2-\Delta+J\right)\left(-1+\Delta+J\right) & 0\\
0 & -\left(2-\Delta+J\right)\left(-1+\Delta+J\right)
\end{array}\right) & m,k=3,4
\end{cases}\label{eq:llll}
\end{equation}

then
\begin{align}
0 & =\det\left(\begin{array}{cc}
\lambda+\left(3-\Delta+J\right)\left(-\Delta+J\right) & -4J\left(\Delta-1\right)\\
-\left(3-\Delta+J\right)\left(1-\Delta+J\right) & \ \lambda-\left(2-\Delta+J\right)\left(1-\Delta+J\right)
\end{array}\right)\nonumber \\
 & =\left(\lambda+\left(3-\Delta+J\right)\left(-\Delta+J\right)\right)\left(\lambda-\left(2-\Delta+J\right)\left(1-\Delta+J\right)\right)-4J\left(\Delta-1\right)\left(3-\Delta+J\right)\left(1-\Delta+J\right)\nonumber \\
 & =\lambda^{2}-\left[\left(2-\Delta+J\right)\left(1-\Delta+J\right)-\left(3-\Delta+J\right)\left(-\Delta+J\right)\right]\lambda+\nonumber \\
 & \ -\left(3-\Delta+J\right)\left(1-\Delta+J\right)\left[\left(-\Delta+J\right)\left(2-\Delta+J\right)+4J\left(\Delta-1\right)\right]\nonumber \\
 & =\lambda^{2}-2\lambda-\left(3-\Delta+J\right)\left(1-\Delta+J\right)\left(\Delta+J\right)\left(J+\Delta-2\right),\\
\lambda_{\pm} & =1\pm\sqrt{1+\left(3-\Delta+J\right)\left(1-\Delta+J\right)\left(\Delta+J\right)\left(-2+\Delta+J\right)}.
\end{align}
Our action then has 4 eigenvalues 
\begin{equation}
\lambda_{\Delta,J}=\text{diag}\begin{pmatrix}1+\sqrt{1+\left(3-\Delta+J\right)\left(1-\Delta+J\right)\left(\Delta+J\right)\left(-2+\Delta+J\right)}\\
1-\sqrt{1+\left(3-\Delta+J\right)\left(1-\Delta+J\right)\left(\Delta+J\right)\left(-2+\Delta+J\right)}\\
\left(2-\Delta+J\right)\left(-1+\Delta+J\right)\\
-\left(2-\Delta+J\right)\left(-1+\Delta+J\right)
\end{pmatrix}.\label{eq:lambdaforreal}
\end{equation}
Let's try to express these in terms of the AdS masses. We can see
that
\begin{align}
-\left(2-\Delta+J\right)\left(-1+\Delta+J\right) & =M_{\Delta,J}^{2}-M_{J+2,J}^{2},\label{eq:mm1}\\
\left(2-\Delta+J\right)\left(-1+\Delta+J\right) & =M_{J+2,J}^{2}-M_{\Delta,J}^{2},\label{eq:mm2}\\
\left(3-\Delta+J\right)\left(1-\Delta+J\right)\left(\Delta+J\right)\left(-2+\Delta+J\right) & =\left(M_{\Delta,J}^{2}-M_{3+J,J}^{2}\right)\left(M_{\Delta,J}^{2}-M_{1+J,J}^{2}\right).
\end{align}

We can see here the physical degrees of freedom which are the zeros
of the quadratic action. For $J=0$ we have the parity odd operator
$\psi\psi$ with $\Delta=J+2=2$. Recall that for $J=0$ the fourth
structure does not exist, so this is matched with the third structure.
Recall also that for odd $J$'s we only have the fourth structure,
while for even $J$'s we have the other three. For $J\geq1$ we have
the parity even operators $\psi\gamma^{\mu}\d_{\mu}\d_{\mu_{1}}\cdots\d_{\mu_{J-1}}\psi$
which have $\Delta=J+1$ for which the second eigenvalue is zero.
Another zero mode exists for $\Delta=J+3$, which together with $\Delta=J+1$
both also exist in the scalar theory. We know from the scalar theory
that while $\Delta=J+1$ is a physical state with a positive norm,
the $\Delta=J+3$ is a negative norm state and is not physical.

Our CFT has four bilocal fields for every spin $J$, and we expect
that these will be related to bulk fields of spin $J$. To do so,
we follow the derivation of the mapping of scalar fields, given in
section 1. Given $G_{\Delta,J}\left(X,Y;S,Z\right)$, the basis of
bulk-to-boundary propagators of massive fields on $AdS$ with spin
$J$ (\ref{eq:bulk2bdry}), one can write the expansion 
\begin{equation}
\Phi_{J}^{m}\left(X,S\right)=\intop_{P.S.}\frac{d\Delta}{2\pi i}\intop\frac{dY}{J!\left(\frac{d}{2}-1\right)_{J}}C_{\Delta,J}^{bulk,m}\left(Y,D_{Z}\right)G_{\Delta,J}\left(X,Y;S,Z\right).\label{eq:Phij}
\end{equation}

One can invert this relation by writing a completeness relation for
our basis. Such a relation is given in \cite{key-22} by defining
the AdS harmonic function 
\begin{equation}
\Omega_{\Delta,J}\left(X_{1},X_{2},S_{1},S_{2}\right)=\intop\frac{dY}{J!\left(\frac{d}{2}-1\right)_{J}}\frac{G_{\Delta,J}\left(X_{1},Y,S_{1},D_{Z}\right)G_{\tilde{\Delta},J}\left(X_{2},Y,S_{3},Z\right)}{N_{\Delta,J}}
\end{equation}
where 
\begin{align}
N_{\Delta,J} & =\frac{2\pi\Gamma\left(\Delta-\frac{d}{2}\right)\Gamma\left(\tilde{\Delta}-\frac{d}{2}\right)}{\Gamma\left(\Delta-1\right)\left(\Delta+J-1\right)\Gamma\left(\tilde{\Delta}-1\right)\left(\tilde{\Delta}+J-1\right)}.
\end{align}
And the completeness relation is given by 
\begin{align}
\delta\left(X_{1},X_{2}\right)\left\langle S_{12}\right\rangle ^{J} & =\intop_{\gamma_{J}}\frac{d\Delta}{2\pi i}\Omega_{\Delta,J}\left(X_{1},X_{2},S_{1},S_{2}\right)+\nonumber \\
 & \ +\sum_{\ell=1}^{J}\left(S_{1}\cdot\d_{X_{1}}\right)^{\ell}\left(S_{2}\cdot\d_{X_{2}}\right)^{\ell}\intop_{\gamma_{J}}\frac{d\Delta}{2\pi i}A_{\Delta,j,\ell}\Omega_{\Delta,J-\ell}\left(X_{1},X_{2},S_{1},S_{2}\right),\\
A_{\Delta,j,\ell} & =\frac{2^{\ell}\left(J-\ell+1\right)_{\ell}\left(\frac{d}{2}+J-\ell-\frac{1}{2}\right)_{\ell}}{\ell!\left(d+2J-2\ell-1\right)_{\ell}\left(d+J-\ell-\Delta\right)_{\ell}\left(\Delta+J-\ell\right)_{\ell}}.
\end{align}
When acting on traceless transverse $\left(TT\right)$ functions only,
as is our case, the second term can be discarded and we are left with
just 
\begin{equation}
\delta^{\left(TT\right)}\left(X_{1},X_{2}\right)\left\langle S_{12}\right\rangle ^{J}=\intop_{\gamma_{J}}\frac{d\Delta}{2\pi i}\Omega_{\Delta,J}\left(X_{1},X_{2},S_{1},S_{2}\right).
\end{equation}

We can now write 
\begin{equation}
C_{\Delta,J}^{bulk,m}\left(Y,Z\right)=\frac{1}{N_{\Delta,J}}\intop\frac{dX}{J!\left(\frac{d}{2}-1\right)_{J}}\Phi_{J}^{m}\left(X,K_{S}\right)G_{\tilde{\Delta},J}\left(X,Y;S,Z\right)
\end{equation}
where $K_{S}$ was defined in (\ref{eq:KS}). we can now find an explicit
mapping between the boundary and bulk theories through 
\begin{equation}
C_{\Delta,J}^{bulk,m}\left(X,S\right)=f_{\Delta,J}^{m}c_{\Delta,J}^{m}\left(X,S\right)
\end{equation}
for some normalization functions $f_{\Delta,J}^{m}$. This normalization
must obey certain consistency conditions. For example, we can get
such a condition by applying the shadow relations to $c_{\Delta,J}^{m},C_{\Delta,J}^{bulk,m}$.
The shadow relation of $c_{\Delta,J}^{m}$ is given by (\ref{eq:shadow}),
and we can write a similar relation for the bulk 
\begin{equation}
G_{\Delta,J}\left(X,Y;S,Z\right)=\frac{1}{S_{B}^{\tilde{\Delta},J}}\intop\frac{dY^{\tg}}{J!\left(\frac{d}{2}-1\right)_{J}}G_{\Delta,J}\left(X,Y^{\tg};S,D_{Z^{\tg}}\right)\left\langle O_{\Delta,J}\left(Y^{\tg},Z^{\tg}\right)O_{\Delta,J}\left(Y,Z\right)\right\rangle \label{eq:Gdj}
\end{equation}
where the bulk shadow coefficients are given by (\ref{eq:bulkshadow}).
We then get 
\begin{equation}
\frac{f_{\Delta,J}^{m}}{f_{\tilde{\Delta},J}^{k}}=\left(\left(S_{\Delta_{\psi},\Delta_{\psi}}^{\tilde{\Delta},J}\right)^{-1}\right)_{\,k}^{m}S_{B}^{\tilde{\Delta},J}.
\end{equation}
Another condition is given by hermiticity of the CFT which implies
\begin{equation}
f_{\Delta,J}^{m*}=\left(-1\right)^{J}f_{\Delta^{*},J}^{m},
\end{equation}
which can be satisfied by requiring $f_{\Delta,J}^{m}$ to be real
analytic functions times a factor of $\left(i\right)^{J}$. We can
then write an explicit mapping from $\rho\left(X_{1},X_{2},S_{1},S_{2}\right)$
to $\Phi_{J}\left(X,W\right)$ by plugging (\ref{eq:cdj}) into (\ref{eq:Phij})
to get 
\begin{align}
\Phi_{J}^{m}\left(X,W\right) & =\frac{1}{2}\sum_{k=1}^{4}\intop_{\gamma_{J}}\frac{d\Delta}{2\pi i}f_{\Delta,J}^{m}\left(\MB_{\Delta,J}^{-1}\right)_{\,k}^{m}\intop\frac{dY}{J!\left(\frac{d}{2}-1\right)_{J}}G_{\Delta,J}\left(X,Y;W,D_{Z}\right)\cross\nonumber \\
 & \ \cross\left\langle O_{\tilde{\Delta}_{\psi}}\left(X_{1},D_{S_{1}}\right)O_{\tilde{\Delta}_{\psi}}\left(X_{2},D_{S_{2}}\right)O_{\tilde{\Delta},J}\left(Y,Z\right)\right\rangle ^{k}\rho\left(X_{1},X_{2},S_{1},S_{2}\right).
\end{align}

From here on we will write the summation over structures explicitly
to avoid confusion. This is an explicit mapping between the CFT degrees
of freedom to the bulk degrees of freedom. Our last step would be
to write the bulk quadratic action. First, we write the bilocal quadratic
action (\ref{eq:squad}) Using embedding space coordinates, restoring
spinor indices 
\begin{equation}
S^{\left(2\right)}=\left(4\pi\right)^{2}\sum_{m,n,k=1}^{4}\sum_{J=0}^{\infty}\intop_{\gamma_{J}}\frac{d\Delta}{2\pi i}\intop\frac{dY}{J!\left(\frac{d}{2}-1\right)_{J}}c_{\Delta,J}^{m}\left(Y,D_{S}\right)\MB_{\Delta,J,n}^{m}A_{\Delta,J,k}^{n}c_{\tilde{\Delta},J}^{k}\left(Y,S\right).
\end{equation}
We can now plug in the value of $c_{\Delta,J}^{m}$ in terms of the
bulk fields, to get 
\begin{equation}
S^{\left(2\right)}\left[\Phi_{J}\right]=\left(4\pi\right)^{2}\sum_{m,n,k=1}^{4}\sum_{J=0}^{\infty}\intop\frac{dX_{1}dX_{2}}{\left(J!\left(\frac{d}{2}-1\right)_{J}\right)^{2}}\Phi_{J}^{m}\left(X_{1},K_{S_{1}}\right)\Phi_{J}^{k}\left(X_{2},K_{S_{2}}\right)\intop_{\gamma_{J}}\frac{d\Delta}{2\pi i}\frac{\MB_{\Delta,J,n}^{m}A_{\Delta,J,k}^{n}}{N_{\Delta,J}f_{\Delta,J}^{m},f_{\tilde{\Delta},J}^{k}}\Omega_{\Delta,J}\left(X_{1},X_{2};S_{1},S_{2}\right).
\end{equation}
Now, recall from (\ref{eq:mm1},\ref{eq:mm2}) that for the odd parity
structures ($n,k=3,4$) we can write 
\begin{equation}
\left(4\pi\right)^{2}A_{\Delta,J,k}^{n}=\left(-1\right)^{n}\left(M_{\Delta,J}^{2}-M_{J+2,J}^{2}\right)\delta_{k}^{n},
\end{equation}
recall that 
\begin{equation}
\del_{X_{1}}^{2}G_{\Delta,J}\left(X_{1},X_{2};S_{1},S_{2}\right)=M_{\Delta,J}^{2}G_{\Delta,J}\left(X_{1},X_{2};S_{1},S_{2}\right),
\end{equation}
so we can write for $n=3,4$ 
\begin{equation}
\left(4\pi\right)^{2}\sum_{k=3,4}A_{\Delta,J,k}^{n}\Phi_{J}^{k}\left(X,S\right)=\left(-1\right)^{n}\left(\del_{X}^{2}-M_{J+2,J}^{2}\right)\Phi_{J}^{n}\left(X,S\right).
\end{equation}
It would be nice to have such a differential operator for the even
parity structures as well. That is, a differential operator in embedding
space that would give us the correct eigenvalues when acting on the
even parity structures. Assuming such an operator can be found, we
could write 
\begin{equation}
\left(4\pi\right)^{2}\sum_{k=1}^{4}A_{\Delta,J,k}^{n}\Phi_{J}^{k}\Phi_{J}\left(X,S\right)=\MD_{k}^{n}\Phi_{J}^{k}\left(X,S\right)
\end{equation}
where for $n=3,4$ we have $\MD_{k}^{n}=\left(\del_{X}^{2}-M_{J+2,J}^{2}\right)\delta_{k}^{n}$
and and for $n=1,2$ some yet unknown $\MD_{k}^{n}$. We will then
get 
\begin{equation}
S^{\left(2\right)}\left[\Phi_{J}\right]=\sum_{m,n,k=1}^{4}\sum_{J=0}^{\infty}\intop\frac{dX_{1}dX_{2}}{\left(J!\left(\frac{d}{2}-1\right)_{J}\right)^{2}}\Phi_{J}^{m}\left(X_{1},K_{S_{1}}\right)\MD_{k}^{n}\Phi_{J}^{k}\left(X_{2},K_{S_{2}}\right)\intop_{\gamma_{J}}\frac{d\Delta}{2\pi i}\frac{\MB_{\Delta,J,n}^{m}}{N_{\Delta,J}f_{\Delta,J}^{m},f_{\tilde{\Delta},J}^{n}}\Omega_{\Delta,J}\left(X_{1},X_{2};S_{1},S_{2}\right).
\end{equation}
For generic $f_{\Delta,J}^{m}$, this quadratic action is non-local.
However, we can choose $f_{\Delta,J}^{m,\text{local}}$ such that
\begin{equation}
\frac{\MB_{\Delta,J,n}^{m}}{N_{\Delta,J}f_{\Delta,J}^{m},f_{\tilde{\Delta},J}^{n}}=\delta_{n}^{m},
\end{equation}
Which, from bulk completeness, would give us the local quadratic action
\begin{equation}
S_{\text{local}}^{\left(2\right)}\left[\Phi_{J}\right]=\sum_{m=1}^{4}\sum_{J=0}^{\infty}\intop\frac{dX}{J!\left(\frac{d}{2}-1\right)_{J}}\Phi_{J}^{m}\left(X,K_{S}\right)\MD_{k}^{n}\Phi_{J}^{k}\left(X,S\right).\label{eq:slocal}
\end{equation}

\subsection{Finding a suitable $\protect\MD_{k}^{1,2}$ operator}

The translation of the first two rows of (\ref{eq:llll}) to bulk
operators is not straightforward, as they are not simply functions
of $M_{\Delta,J}^{2}$ which we can map to the bulk Laplacian. However
their eigenvalues are such functions, so we try to first change the
basis so that it takes a simpler form. We wish to write a $2\cross2$
matrix $\lambda$ such that 
\begin{align}
\det\lambda & =-\left(M_{\Delta,J}^{2}-M_{3+J,J}^{2}\right)\left(M_{\Delta,J}^{2}-M_{1+J,J}^{2}\right),\\
\Tr\lambda & =2.
\end{align}

Take the ansatz 
\begin{equation}
\lambda=\begin{pmatrix}1+\frac{1}{\sqrt{2}}\left(M_{\Delta,J}^{2}-N_{1}\right) & 1+\frac{i}{\sqrt{2}}\left(M_{\Delta,J}^{2}-N_{2}\right)\\
1-\frac{i}{\sqrt{2}}\left(M_{\Delta,J}^{2}-N_{2}\right) & 1-\frac{1}{\sqrt{2}}\left(M_{\Delta,J}^{2}-N_{1}\right)
\end{pmatrix}.
\end{equation}

It is clear that the trace condition is satisfied. The determinant
is 
\begin{align}
\det\lambda & =-\frac{1}{2}\left(M_{\Delta,J}^{2}-N_{1}\right)^{2}-\frac{1}{2}\left(M_{\Delta,J}^{2}-N_{2}\right)^{2}\nonumber \\
 & =-M_{\Delta,J}^{4}+M_{\Delta,J}^{2}\left(N_{1}+N_{2}\right)-\frac{1}{2}\left(N_{1}^{2}+N_{2}^{2}\right)\nonumber \\
 & =-M_{\Delta,J}^{4}+M_{\Delta,J}^{2}\left(M_{1+J,J}^{2}+M_{3+J,J}^{2}\right)-M_{3+J,J}^{2}M_{1+J,J}^{2},
\end{align}
so we have 
\begin{align}
N_{1}^{2}-N_{1}\left(M_{1+J,J}^{2}+M_{3+J,J}^{2}\right)+\frac{1}{2}\left(M_{1+J,J}^{2}-M_{3+J,J}^{2}\right)^{2} & =0,
\end{align}
which gives 
\begin{align}
N_{1} & =\frac{1}{2}\left(M_{1+J,J}^{2}+M_{3+J,J}^{2}\right)+\frac{i}{2}\left(M_{1+J,J}^{2}-M_{3+J,J}^{2}\right),\\
N_{2} & =\frac{1}{2}\left(M_{1+J,J}^{2}+M_{3+J,J}^{2}\right)-\frac{i}{2}\left(M_{1+J,J}^{2}-M_{3+J,J}^{2}\right)
\end{align}

or equivalently the other way around. This gives us an explicit local
operator in the bulk, just by replacing $M_{\Delta,J}^{2}$ with $\del_{X_{1}}^{2}$.
The price we pay is that the operator is complex, and the meaning
of that is unclear, however we know it will not cause problems because
it is equivalent to the original CFT action which makes sense.

We make another attempt of finding such an operator, by turning our
attention back to the CFT. We wish to find an embedding space operator,
which when acting on even parity structures will give us the coefficients
matrix which appears in the first case of (\ref{eq:llll}). As our
coordinate space action is given by $S_{qa}=x_{12}^{2}\cancel{\d}_{1\beta}^{\gamma}\cancel{\d}_{2\alpha}^{\delta}$,
a natural candidate would be 
\begin{equation}
S_{emb}=-S_{1K}S_{2L}\Gamma_{AJ}^{K}\Gamma_{BI}^{L}\frac{\d}{\d S_{1I}}\frac{\d}{\d S_{2J}}X_{12}\d_{1}^{A}\d_{2}^{B}.
\end{equation}

Let's see how this acts on our four 3-pt. structures. On the first
structure we get 
\begin{equation}
S_{emb}\left\langle \Psi_{1}\Psi_{2}O^{\Delta,J}\right\rangle ^{1}=\left(d-\Delta+J\right)\left(J-\Delta\right)\left\langle \Psi_{1}\Psi_{2}O^{\Delta,J}\right\rangle ^{1}-4J\left(\Delta-1\right)\left\langle \Psi_{1}\Psi_{2}O^{\Delta,J}\right\rangle ^{2}.
\end{equation}

Recall that we had 
\begin{equation}
s_{qa}\left\langle \psi\psi O^{\Delta,J}\right\rangle ^{1}=\left(d-\Delta+J\right)\left(J-\Delta\right)\left\langle \psi\psi O^{\Delta,J}\right\rangle ^{1}-4J\left(\Delta-1\right)\left\langle \psi\psi O^{\Delta,J}\right\rangle ^{2},
\end{equation}

so this indeed seems to match what we expect from $S_{emb}$. For
the second structure 
\begin{align}
S_{emb}\left\langle \Psi_{1}\Psi_{2}O^{\Delta,J}\right\rangle ^{2} & =\left(1-\Delta+J\right)\left(2-\Delta+J\right)\left\langle \Psi_{1}\Psi_{2}O^{\Delta,J}\right\rangle ^{2}+\left(1-\Delta+J\right)\left(3-\Delta+J\right)\left\langle \Psi_{1}\Psi_{2}O^{\Delta,J}\right\rangle ^{1}
\end{align}
also, as expected from $S_{qa}$. For the third and fourth structures
we get
\begin{equation}
S_{emb}\left\langle \Psi_{1}\Psi_{2}O^{\Delta,J}\right\rangle ^{3,4}=0.
\end{equation}

The fact that this operator maps the space spanned by the 3-pt. function
into a subspace of it, tells us that it is well-defined in embedding
space, i.e. that it preserves the ideal generated by the nullness
and traverseness conditions of the embedding space variables. We conclude
that there is indeed an embedding space differential operator which
can act on the 3-pt. structures and reproduce the correct coefficients
$\lambda_{\Delta,J}$ on the even parity structures, and in addition
contributes nothing when acts on the odd parity structures. We can
hope that this operator can take a local form in AdS, by acting with
it on the bulk to boundary propagator basis and restricting the embedding
space coordinates to the AdS hyperboloid. This will be done in future
research.

\section{Discussion}

To summarize, this paper achieves an explicit mapping between the
theory of free Majorana fermions in 3 dimensions, and a high-spin
gravity theory in $AdS_{4}$, which we believe should be equivalent
to type II Vasiliev gravity, since the latter was also shown to lead
to the same correlation functions as the free fermion theory (at least
in some cases that were tested, see \cite{key-24}). This mapping
gives a quantization of a notoriously difficult theory of gravity.
There are however many questions that remain open and allow for much
future research: 
\begin{enumerate}
\item What is the explicit form of the bulk quadratic action for parity
even operators? 
\item Can this be generalized to other spacetime dimensions and fermion
representations? It would appear that this is the case as we have
mainly used 3D Majorana fermions for simplicity and not for any specific
properties they have.
\item How do the interactions of the bilocal theory map to the bulk and
what are the Feynman rules?
\item What happens when we add deformations (mass, double trace, etc.) to
the theory?
\item Can we show that the theory we get is equivalent to an off-shell version
of Vasiliev's theory (which is so far known only as on-shell equations
of motion)?
\item Can these results be generalized to finite $N$, volume and temperature
\end{enumerate}
The last point was achieved recently by Aharony et al. \cite{key-7}
for the bosonic case, and it stands to reason that it can similarly
be done for the fermionic case.

\subsection*{Acknowledgments}

I thank Ofer Aharony for advising me during this work. This work was
supported in part by an Israel Science Foundation (ISF) center for
excellence grant (grant number 2289/18), by ISF grant no. 2159/22,
by Simons Foundation grant 994296 (Simons Collaboration on Confinement
and QCD Strings), by grant no. 2018068 from the United States-Israel
Binational Science Foundation (BSF), by the Minerva foundation with
funding from the Federal German Ministry for Education and Research,
by the German Research Foundation through a German-Israeli Project
Cooperation (DIP) grant ``Holography and the Swampland'', and by
a research grant from Martin Eisenstein.

\appendix

\section{Embedding space formalism}

In this work we make use of the embedding space formalism, following
the notations of \cite{key-12}. The $d$ dimensional Lorentzian conformal
group is isomorphic to $SO\left(d,2\right)$, which is the rotation
group of $d+2$ dimensional flat spacetime with metric $\eta_{AB}=\eta^{AB}=\text{diag}\left(-1,1,\cdots,1,-1\right)$,
$A,B=0,\cdots,d+1$. The $SO\left(d,2\right)$ generators $J^{AB}$
can be related to the $d$ dimensional conformal group generators
through ($\mu,\nu=0,\dots,d-1$)
\begin{align}
D & =-J^{d,d+1},\ P^{\mu}=J^{d,\mu}+J^{d+1,\mu},\\
K^{\mu} & =-J^{d,\mu}+J^{d+1,\mu},\ M^{\mu\nu}=J^{\mu\nu}.
\end{align}
We denote the embedding space coordinates $X^{A}$, and note that
our $d$ dimensional spacetime is embedded as the projective null
cone, which are the points that satisfy $X\cdot X=0$ and are identified
up to the rescaling $X^{A}\sim\lambda X^{A}$. In light-cone coordinates
$X=\left(X^{\mu},X^{+},X^{-}\right)$ where $X^{\pm}=X^{d+1}\pm X^{d}$
we can parameterize our light cone as 
\begin{equation}
x^{\mu}=\frac{X^{\mu}}{X^{+}},\ X=X^{+}\left(x^{\mu},1,x^{2}\right)
\end{equation}
where $x^{2}=\eta_{\mu\nu}x^{\mu}x^{\nu}$. Without loss of generality
we can take $X^{+}=1$. We define 
\begin{equation}
X_{ij}\defined-2X_{i}\cdot X_{j}=-2x_{i}\cdot x_{j}+x_{i}^{2}+x_{j}^{2}=\left(x_{i}-x_{j}\right)^{2}\defined x_{ij}^{2}.
\end{equation}

We will be interested in spinor fields that are primaries of the conformal
group. We embed primary spinor fields by noting how the conformal
group acts on them 
\begin{align}
i\left[M^{\mu\nu},\psi^{\alpha}\left(x\right)\right] & =\left(x^{\nu}\d^{\mu}-x^{\mu}\d^{\nu}\right)\psi^{\alpha}\left(x\right)-i\left(\MM^{\mu\nu}\right)_{\,\beta}^{\alpha}\psi^{\beta}\left(x\right),\nonumber \\
i\left[P^{\mu},\psi^{\alpha}\left(x\right)\right] & =-\d^{\mu}\psi^{\alpha}\left(x\right),\nonumber \\
i\left[K^{\mu},\psi^{\alpha}\left(x\right)\right] & =\left(2x^{\mu}x^{\nu}\d_{\nu}-x^{2}\d^{\mu}+2\Delta_{\psi}x^{\mu}\right)\psi^{\alpha}\left(x\right)+2ix_{\nu}\left(\MM^{\nu\mu}\right)_{\,\beta}^{\alpha}\psi^{\beta}\left(x\right),\nonumber \\
i\left[D,\psi^{\alpha}\left(x\right)\right] & =\left(x^{\mu}\d_{\mu}+\Delta_{\psi}\right)\psi^{\alpha}\left(x\right).\label{eq:confgen}
\end{align}
Where $\MM^{\mu\nu}=-\frac{i}{4}\left[\gamma^{\mu},\gamma^{\nu}\right]$,
and $\alpha$,$\beta$ are spinor indices, which can be raised and
lowered with the symplectic form $\Omega^{\alpha\beta}=\Omega_{\alpha\beta}$
which preserves $\Omega M^{\mu\nu}+\left(M^{\mu\nu}\right)^{T}\Omega=0$.
To keep track of spinor indices, we will introduce a set of auxiliary
commuting variables $s_{\alpha},\overline{s}^{\dot{\alpha}}$ such
that 
\begin{align}
\psi\left(x,s\right) & \defined s_{\alpha}\psi^{\alpha}\left(x\right),\\
\overline{\psi}\left(x,\overline{s}\right) & \defined\overline{s}^{\dot{\alpha}}\overline{\psi}_{\dot{\alpha}}\left(x\right).
\end{align}

In general we can write operators $\MO$ in the $\left(\ell,\overline{\ell}\right)$
spin representations (note that these representations only exist in
certain spacetime dimensions, but we will use them for general dimensions
and set $\ell=\overline{\ell},\ \alpha=\dot{\alpha}$ in spacetime
dimensions where there is only a single spin quantum number) 
\begin{align}
\MO\left(x,s,\overline{s}\right) & =s_{\alpha_{1}}\cdots s_{\alpha_{\ell}}\overline{s}^{\dot{\alpha}_{1}}\cdots\overline{s}^{\dot{\alpha}_{\overline{\ell}}}\MO_{\dot{\alpha}_{1}\cdots\dot{\alpha}_{\overline{\ell}}}^{\alpha_{1}\dots\alpha_{\ell}}\left(x\right),\\
\MO_{\dot{\alpha}_{1}\cdots\dot{\alpha}_{\overline{\ell}}}^{\alpha_{1}\dots\alpha_{\ell}}\left(x\right) & =\frac{1}{\ell!\overline{\ell}!}\frac{\d^{\ell}}{\d s_{\alpha_{1}}\cdots\d s_{\alpha_{\ell}}}\frac{\d^{\overline{\ell}}}{\d s_{\dot{\alpha}_{1}}\cdots\d s_{\dot{\alpha}_{\overline{\ell}}}}\MO\left(x,s,\overline{s}\right).
\end{align}
Traceless symmetric tensors $\left(\ell,\ell\right)$ will have integer
spin and can be written as 
\begin{equation}
\MO^{\mu_{1}\cdots\mu_{\ell}}\left(x\right)=\frac{\left(-1\right)^{\ell}}{2^{\ell}}\gamma_{\alpha_{1}}^{\mu_{1}\dot{\alpha}_{1}}\cdots\gamma_{\alpha_{\ell}}^{\mu_{\ell}\dot{\alpha}_{\ell}}\MO_{\dot{\alpha}_{1}\cdots\dot{\alpha}_{\ell}}^{\alpha_{1}\dots\alpha_{\ell}}\left(x\right).
\end{equation}

Note that for real representations (e.g. Majorana fermions) we will
have $s_{\alpha}=\overline{s}^{\dot{\alpha}}$. In this work we use
only these representations, so we assume this from now on. Operators
in embedding space are identified with their physical space counterparts
(up to a power of $X^{+}$ which we will ignore) 
\begin{equation}
\MO\left(X,S\right)=\MO\left(x,s\right)
\end{equation}

with 
\begin{equation}
S_{I}=\begin{pmatrix}s_{\alpha}\\
-x^{\alpha\beta}s_{\beta}
\end{pmatrix},\ S^{I}=\begin{pmatrix}s_{\beta}x^{\beta\alpha}\\
s_{\alpha}
\end{pmatrix}
\end{equation}
where $x^{\alpha\beta}=x^{\mu}\gamma_{\mu}^{\alpha\beta}$. $S_{I}$
satisfies the transversality condition $S_{I}X_{\,J}^{I}=0$ where
$X_{\,J}^{I}=X^{A}\Gamma_{AJ}^{I}$ and $\Gamma_{A}$ are the embedding
space Dirac matrices which satisfy the Clifford algebra $\left\{ \Gamma_{A},\Gamma_{B}\right\} =2\eta_{AB}$.
We will define for 2 spin variables $S_{i},S_{j}$ and a set of embedding
space coordinates $X_{1},\dots,X_{k}$ 
\begin{equation}
\left\langle S_{i}X_{1}\cdots X_{k}S_{j}\right\rangle \defined S_{iI}X_{1J}^{I}\cdots X_{kL}^{K}\Omega^{LM}S_{jM}
\end{equation}
where $\Omega^{LM}$ is the embedding space symplectic form. Note
that this satisfies 
\begin{align}
\left\langle S_{i}S_{j}\right\rangle  & =-\left\langle S_{j}S_{i}\right\rangle ,\\
\left\langle S_{i}S_{i}\right\rangle  & =0,\\
\left\langle S_{k}X_{i}S_{k}\right\rangle  & =0,\\
\left\langle S_{k}X_{i}X_{j}S_{k}\right\rangle  & =-\left\langle S_{k}X_{j}X_{i}S_{k}\right\rangle ,
\end{align}
as well as the Fierz identities (here we assume transversality $S_{iI}X_{iJ}^{I}=0$)
\begin{align}
\left\langle S_{1}X_{2}S_{3}\right\rangle \left\langle S_{2}X_{1}S_{3}\right\rangle  & =-\left\langle S_{1}S_{2}\right\rangle \left\langle S_{3}X_{1}X_{2}S_{3}\right\rangle +2\left\langle S_{2}S_{3}\right\rangle \left\langle S_{1}S_{3}\right\rangle X_{1}\cdot X_{2},\\
\left\langle S_{1}X_{3}S_{2}\right\rangle \left\langle S_{3}X_{1}X_{2}S_{3}\right\rangle  & =-2\left\langle S_{1}S_{3}\right\rangle \left\langle S_{2}X_{1}S_{3}\right\rangle X_{2}\cdot X_{3}-2\left\langle S_{2}S_{3}\right\rangle \left\langle S_{1}X_{2}S_{3}\right\rangle X_{1}\cdot X_{3}.
\end{align}
Included here are some more useful identities, mostly used in section
2.5. 
\begin{align}
\left\langle S_{1}\Gamma_{A}\Gamma_{B}S_{2}\right\rangle \left\langle S_{3}X_{1}\Gamma^{B}S_{3}\right\rangle \left\langle S_{3}\Gamma^{A}X_{2}S_{3}\right\rangle  & =4\left\langle S_{3}X_{1}X_{2}S_{3}\right\rangle \left\langle S_{1}S_{3}\right\rangle \left\langle S_{2}S_{3}\right\rangle ,\\
\left\langle S_{1}\Gamma_{A}\Gamma_{B}S_{2}\right\rangle \left\langle S_{3}\Gamma^{A}\Gamma^{B}S_{3}\right\rangle  & =8\left\langle S_{1}S_{3}\right\rangle \left\langle S_{2}S_{3}\right\rangle ,\\
\left\langle S_{1}\Gamma_{A}\Gamma_{B}S_{2}\right\rangle \left(\left\langle S_{3}\Gamma^{A}X_{2}S_{3}\right\rangle X_{1}^{B}+X_{2}^{A}\left\langle S_{3}X_{1}\Gamma^{B}S_{3}\right\rangle \right) & =4\left\langle S_{3}X_{1}X_{2}S_{3}\right\rangle \left\langle S_{1}S_{2}\right\rangle ,\\
\left\langle S_{1}\Gamma_{A}\Gamma_{B}S_{2}\right\rangle \left\langle S_{3}\Gamma^{A}X_{2}S_{3}\right\rangle X_{3}^{B} & =-2\left\langle S_{1}S_{3}\right\rangle \left\langle S_{2}S_{3}\right\rangle X_{23},\\
\left\langle S_{1}\Gamma_{A}\Gamma_{B}S_{2}\right\rangle \left\langle S_{3}X_{1}\Gamma^{B}S_{3}\right\rangle X_{3}^{A} & =-2\left\langle S_{1}S_{3}\right\rangle \left\langle S_{2}S_{3}\right\rangle X_{31},\\
\left\langle S_{1}\Gamma_{A}\Gamma_{B}S_{2}\right\rangle \eta^{AB} & =D\left\langle S_{1}S_{2}\right\rangle ,\\
\left\langle S_{1}\Gamma_{A}\Gamma_{B}S_{2}\right\rangle X_{3}^{A}X_{1}^{B} & =-\left\langle S_{1}S_{2}\right\rangle X_{31},\\
\left\langle S_{1}\Gamma_{A}\Gamma_{B}S_{2}\right\rangle X_{3}^{A}X_{3}^{B} & =0,\\
\left\langle S_{1}\Gamma_{A}\Gamma_{B}S_{2}\right\rangle X_{2}^{A}X_{1}^{B} & =-\left\langle S_{1}S_{2}\right\rangle X_{12},\\
\left\langle S_{1}\Gamma_{A}\Gamma_{B}S_{2}\right\rangle \left(X_{2}^{A}X_{3}^{B}+X_{2}^{B}X_{3}^{A}\right) & =-\left\langle S_{1}S_{2}\right\rangle X_{23},\\
\left\langle S_{1}\Gamma_{A}S_{3}\right\rangle \left\langle S_{2}\Gamma_{B}S_{3}\right\rangle \eta^{AB} & =\left\langle S_{1}S_{3}\right\rangle \left\langle S_{2}S_{3}\right\rangle ,\\
\left\langle S_{1}\Gamma_{A}S_{3}\right\rangle \left\langle S_{2}\Gamma_{B}S_{3}\right\rangle X_{2}^{A}X_{1}^{B} & =\left\langle S_{1}X_{2}S_{3}\right\rangle \left\langle S_{2}X_{1}S_{3}\right\rangle .
\end{align}
Another useful identity for $D=5$ Dirac matrices is
\begin{equation}
\Gamma_{\ \,J}^{AI}\Gamma_{AL}^{K}=2\delta_{\,J}^{K}\delta_{\,L}^{I}-\delta_{\,J}^{I}\delta_{\,L}^{K}+2\Omega^{IK}\Omega_{JL}.
\end{equation}

\section{Useful properties of the fermion 3-pt. functions}

\subsection{Calculation of 3-pt. function pairings}

This appendix contains properties of the fermion 3-pt. functions used
throughout the paper. We work with these functions by noting that
they can be expanded using the scalar 3-pt. functions when operated
on by weight shifting operators. The expansion is given by \cite{key-11}
\begin{equation}
\left\langle \psi_{1}^{\Delta_{1}}\psi_{2}^{\Delta_{2}}O^{\Delta,J}\right\rangle ^{m}=\sum_{p,q=\pm1}\kappa_{3,pq}^{m}\left(\psi_{1}^{\Delta_{1}}\psi_{2}^{\Delta_{2}}O^{\Delta,J}\right)\MD_{1}^{-p,+}\MD_{2}^{-q,+}\left\langle \phi_{1}^{\Delta_{1}+p/2}\phi_{2}^{\Delta_{2}+q/2}O^{\Delta,J}\right\rangle 
\end{equation}
 where $\kappa_{3,pq}^{m}\left(\psi_{1}^{\Delta_{1}}\psi_{2}^{\Delta_{2}}O^{\Delta,J}\right)$
are the expansion coefficients and can be expressed as an invertible
$4\cross4$ matrix, and $\MD_{i}^{pq}$ are the weight shifting operators
acting on the $i$th field at $\left(X_{i},S_{i}\right)$ and are
given in embedding space by 
\begin{align}
\MD_{I}^{-+} & =S_{I},\\
\MD_{I}^{++} & =-2\left(\Delta-1\right)S_{J}\left(\d_{X}\right)_{\,I}^{J}-S_{I}S_{J}\left(\d_{X}\right)_{\,K}^{J}\frac{\d}{\d S_{K}},\\
\MD_{I}^{--} & =X_{IJ}\frac{\d}{\d S_{J}},\\
\MD_{I}^{+-} & =4\left(1+\ell-\Delta\right)\left(\Delta-1\right)\frac{\d}{\d S^{I}}+2\left(1+\ell-\Delta\right)X_{IJ}\left(\d_{X}\right)_{\,K}^{J}\frac{\d}{\d S_{K}}-S_{I}\frac{\d}{\d S_{J}}X_{JK}\left(\d_{X}\right)_{\,L}^{K}\frac{\d}{\d S_{L}},
\end{align}
where $A,B,C,D$ are embedding space spinor indices, and $\left(\d_{X}\right)_{\,J}^{I}=\Gamma_{AJ}^{I}\frac{\d}{\d X_{A}}$.
We can calculate the coefficients $\kappa_{3}$ using 

\begin{align}
\MD_{1}^{-+}\MD_{2}^{-+}\left\langle \phi_{1}^{\Delta_{\psi}+1/2}\phi_{2}^{\Delta_{\psi}+1/2}O^{\Delta,J}\right\rangle  & =\left\langle \psi_{1}\psi_{2}O^{\Delta,J}\right\rangle ^{1},\\
\MD_{1}^{++}\MD_{2}^{-+}\left\langle \phi_{1}^{\Delta_{\psi}-1/2}\phi_{2}^{\Delta_{\psi}+1/2}O^{\Delta,J}\right\rangle  & =2\left(\Delta_{\psi}-3/2\right)\left(\left(\Delta+2J-1\right)\left\langle \psi_{1}\psi_{2}O^{\Delta,J}\right\rangle ^{3}+J\left\langle \psi_{1}\psi_{2}O^{\Delta,J}\right\rangle ^{4}\right),\\
\MD_{1}^{-+}\MD_{2}^{++}\left\langle \phi_{1}^{\Delta_{\psi}+1/2}\phi_{2}^{\Delta_{\psi}-1/2}O^{\Delta,J}\right\rangle  & =2\left(\Delta_{\psi}-3/2\right)\left(\left(\Delta+2J-1\right)\left\langle \psi_{1}\psi_{2}O^{\Delta,J}\right\rangle ^{3}-J\left\langle \psi_{1}\psi_{2}O^{\Delta,J}\right\rangle ^{4}\right),\\
\MD_{1}^{++}\MD_{2}^{++}\left\langle \phi_{1}^{\Delta_{\psi}-1/2}\phi_{2}^{\Delta_{\psi}-1/2}O^{\Delta,J}\right\rangle  & =-4\left(\Delta_{\psi}-3/2\right)^{2}\paren{4J\left(J-3\right)\left\langle \psi_{1}\psi_{2}O^{\Delta,J}\right\rangle ^{2}+}\nonumber \\
 & \quad\thesis{+\left(2\Delta_{\psi}-\Delta+J+1\right)\left(2\Delta_{\psi}-\Delta+J-2\right)\left\langle \psi_{1}\psi_{2}O^{\Delta,J}\right\rangle ^{1}}.
\end{align}
This expansion can be used to calculate pairings of 3-pt. functions,
which are defined by These obey the property of integration by parts
of the weight shifting operators 
\begin{align}
\left(\MD_{\Delta,J}^{pq}\MO,\tilde{\MO}^{\tg}\right) & =\left(\MO,\left(\MD_{\Delta,J}^{pq}\right)^{*}\tilde{\MO}^{\tg}\right)\nonumber \\
 & =\left(\MO,\zeta_{J}^{pq}\MD_{\tilde{\Delta}-\frac{p}{2},J+\frac{q}{2}}^{p,-q}\tilde{\MO}^{\tg}\right)
\end{align}
where 
\begin{align}
\zeta_{J}^{--} & =-\zeta_{J}^{+-}=-2J,\\
\zeta_{J}^{-+} & =-\zeta_{J}^{++}=\frac{1}{2J+1}.
\end{align}
And we get 
\begin{equation}
\left(\left\langle \psi_{1}^{\Delta_{1}}\psi_{2}^{\Delta_{2}}O^{\Delta,J}\right\rangle ^{m},\left\langle \psi_{1}^{\tilde{\Delta}_{1}}\psi_{2}^{\tilde{\Delta}_{2}}O^{\tilde{\Delta},J}\right\rangle ^{n}\right)=\begin{cases}
\frac{1}{8\pi}\begin{pmatrix}-1 & 0\\
0 & 1
\end{pmatrix} & J=0\\
\frac{\left(-1\right)^{J}\Gamma\left(J+1\right)}{8\sqrt{\pi}\Gamma\left(J+\frac{1}{2}\right)}\begin{pmatrix}-1 & 1/2 & 0 & 0\\
1/2 & -\frac{2J+1}{4J} & 0 & 0\\
0 & 0 & 1 & 0\\
0 & 0 & 0 & -\frac{J+1}{J}
\end{pmatrix} & J>0
\end{cases}.
\end{equation}

\subsection{Shadow coefficients for the orthogonality of the 3-pt. functions}

The shadow coefficients that can also calculated using the expansion
method above. This gives (Note that these generally depend on the
difference between the conformal weights of the 2 fermions. We assume
here that the fermions are identical so there is no dependence on
$\Delta_{\psi}$) 
\begin{align}
S_{1}^{1}\left(\psi\psi\left[\MO_{\Delta,J}\right]\right) & =\frac{\pi^{3/2}\left(-1\right)^{J}\Gamma\left(J+\Delta-1\right)\Gamma\left(\frac{J+\tilde{\Delta}}{2}\right)^{2}}{\Gamma\left(\Delta-1\right)\Gamma\left(J+\tilde{\Delta}\right)\Gamma\left(\frac{J+\Delta}{2}\right)^{2}},\\
S_{1}^{2}\left(\psi\psi\left[\MO_{\Delta,J}\right]\right) & =-\frac{\Delta-3/2}{\Delta-1}S_{1}^{1}\left(\psi\psi\left[\MO_{\Delta,J}\right]\right),\\
S_{2}^{2}\left(\psi\psi\left[\MO_{\Delta,J}\right]\right) & =-\frac{\Delta-2}{\Delta-1}S_{1}^{1}\left(\psi\psi\left[\MO_{\Delta,J}\right]\right),\\
S_{3}^{3}\left(\psi\psi\left[\MO_{\Delta,J}\right]\right) & =\frac{\left(\Delta-1\right)\left(\Delta-2\right)^{2}-\left(\Delta-2\right)J\left(J+1\right)}{2\left(2\Delta-3\right)\left(-\Delta+J+2\right)\left(\Delta+J-1\right)}S_{1}^{1}\left(\psi\psi\left[\MO_{\Delta+1,J}\right]\right),\\
S_{4}^{4}\left(\psi\psi\left[\MO_{\Delta,J}\right]\right) & =\frac{-\left(\Delta-2\right)\left(\Delta-1\right)^{2}+\left(\Delta-1\right)J\left(J+1\right)}{2\left(2\Delta-3\right)\left(-\Delta+J+2\right)\left(\Delta+J-1\right)}S_{1}^{1}\left(\psi\psi\left[\MO_{\Delta+1,J}\right]\right),
\end{align}
and the rest vanish. 

\subsection{Shadow coefficients for the shadow transform of the 3-pt. functions}

The shadow coefficients for equation (\ref{eq:shadowtrans}) are

\begin{align}
S_{3}^{1}\left(\left[\psi^{\Delta_{1}}\right]\psi^{\Delta_{2}}\MO^{\Delta,J}\right) & =\frac{i\pi^{3/2}\left(-\Delta+\Delta_{1}+\Delta_{2}-2\right)\Gamma\left(\Delta_{1}-1\right)\Gamma\left(\frac{J+\Delta-\Delta_{1}-\Delta_{2}+2}{2}\right)\Gamma\left(\frac{J-\Delta-\Delta_{1}+\Delta_{2}+3}{2}\right)}{2\Gamma\left(\frac{7}{2}-\Delta_{1}\right)\Gamma\left(\frac{J+\Delta+\Delta_{1}-\Delta_{2}}{2}\right)\Gamma\left(\frac{J-\Delta+\Delta_{1}+\Delta_{2}+1}{2}\right)},\\
S_{3}^{1}\left(\psi^{\Delta_{1}}\left[\psi^{\Delta_{2}}\right]\MO^{\Delta,J}\right) & =S_{3}^{1}\left(\left[\psi^{\Delta_{2}}\right]\psi^{\Delta_{1}}\MO^{\Delta,J}\right),\\
\frac{S_{4}^{1}\left(\left[\psi^{\Delta_{1}}\right]\psi^{\Delta_{2}}\MO^{\Delta,J}\right)}{\frac{J}{\Delta-\Delta_{1}-\Delta_{2}+2}} & =\frac{S_{3}^{2}\left(\left[\psi^{\Delta_{1}}\right]\psi^{\Delta_{2}}\MO^{\Delta,J}\right)}{\frac{-\Delta+\Delta_{1}+\Delta_{2}+J-1}{2\left(\Delta-\Delta_{1}-\Delta_{2}+2\right)}}=\frac{S_{4}^{2}\left(\left[\psi^{\Delta_{1}}\right]\psi^{\Delta_{2}}\MO^{\Delta,J}\right)}{\frac{\Delta-\Delta_{1}-\Delta_{2}-J+1}{2\left(\Delta-\Delta_{1}-\Delta_{2}+2\right)}}=S_{3}^{1}\left(\left[\psi^{\Delta_{1}}\right]\psi^{\Delta_{2}}\MO^{\Delta,J}\right),\\
\frac{S_{1}^{3}\left(\left[\psi^{\Delta_{1}}\right]\psi^{\Delta_{2}}\MO^{\Delta,J}\right)}{\frac{-\Delta-\Delta_{1}+\Delta_{2}+J+2}{\Delta+\Delta_{1}-\Delta_{2}-2}} & =\frac{S_{2}^{3}\left(\left[\psi^{\Delta_{1}}\right]\psi^{\Delta_{2}}\MO^{\Delta,J}\right)}{\frac{2J}{\Delta+\Delta_{1}-\Delta_{2}-2}}=\frac{S_{2}^{4}\left(\left[\psi^{\Delta_{1}}\right]\psi^{\Delta_{2}}\MO^{\Delta,J}\right)}{\frac{-2\Delta-2\Delta_{1}+2\Delta_{2}+2}{\Delta+\Delta_{1}-\Delta_{2}-2}}=S_{3}^{1}\left(\left[\psi^{\Delta_{1}}\right]\psi^{-\Delta_{2}}\MO^{-\Delta,J}\right),\\
S_{1}^{4}\left(\left[\psi^{\Delta_{1}}\right]\psi^{\Delta_{2}}\MO^{\Delta,J}\right) & =S_{1}^{3}\left(\left[\psi^{\Delta_{1}}\right]\psi^{\Delta_{2}}\MO^{\Delta,J}\right),\\
\frac{S_{4}^{1}\left(\psi^{\Delta_{1}}\left[\psi^{\Delta_{2}}\right]\MO^{\Delta,J}\right)}{\frac{-J}{\Delta-\Delta_{1}-\Delta_{2}+2}} & =\frac{S_{3}^{2}\left(\psi^{\Delta_{1}}\left[\psi^{\Delta_{2}}\right]\MO^{\Delta,J}\right)}{\frac{-\Delta+\Delta_{1}+\Delta_{2}+J-1}{2\left(\Delta-\Delta_{1}-\Delta_{2}+2\right)}}=\frac{S_{4}^{2}\left(\psi^{\Delta_{1}}\left[\psi^{\Delta_{2}}\right]\MO^{\Delta,J}\right)}{\frac{-\Delta+\Delta_{1}+\Delta_{2}+J-1}{2\left(\Delta-\Delta_{1}-\Delta_{2}+2\right)}}=S_{3}^{1}\left(\psi^{\Delta_{1}}\left[\psi^{\Delta_{2}}\right]\MO^{\Delta,J}\right),\\
\frac{S_{1}^{3}\left(\psi^{\Delta_{1}}\left[\psi^{\Delta_{2}}\right]\MO^{\Delta,J}\right)}{\frac{-\Delta+\Delta_{1}-\Delta_{2}+J+2}{\Delta-\Delta_{1}+\Delta_{2}-2}} & =\frac{S_{2}^{3}\left(\psi^{\Delta_{1}}\left[\psi^{\Delta_{2}}\right]\MO^{\Delta,J}\right)}{\frac{2J}{\Delta-\Delta_{1}+\Delta_{2}-2}}=\frac{S_{2}^{4}\left(\psi^{\Delta_{1}}\left[\psi^{\Delta_{2}}\right]\MO^{\Delta,J}\right)}{\frac{2\Delta-2\Delta_{1}+2\Delta_{2}-2}{\Delta-\Delta_{1}+\Delta_{2}-2}}=S_{3}^{1}\left(\psi^{-\Delta_{1}}\left[\psi^{\Delta_{2}}\right]\MO^{-\Delta,J}\right),\\
S_{1}^{4}\left(\psi^{\Delta_{1}}\left[\psi^{\Delta_{2}}\right]\MO^{\Delta,J}\right) & =-S_{1}^{3}\left(\psi^{\Delta_{1}}\left[\psi^{\Delta_{2}}\right]\MO^{\Delta,J}\right).
\end{align}
The other coefficients vanish.

\section{Mathematica code}

The following is a Mathematica code which defines the 3-point structures
and acts on them with the operator $S_{qa}$ to give (\ref{eq:sqa1}).

Embedding space $d=5$
\begin{lyxcode}
d~=~5;
\end{lyxcode}
Metric $eta=\eta_{AB}$ and inverse metric $etaa$
\begin{lyxcode}
eta~=~\{\{-1,~0,~0,~0,~0\},~\{0,~1,~0,~0,~0\},~\{0,~0,~1,~0,~0\},

~~~~~~~\{0,~0,~0,~0,~-1/2\},~\{0,~0,~0,~-1/2,~0\}\};

etaa~=~Inverse{[}eta{]};
\end{lyxcode}
Defining the spacetime coordinates $X_{1}^{A},X_{2}^{A},X_{3}^{A}$
\begin{lyxcode}
x1~=~Table{[}X1{[}i{]},~\{i,~0,~d~-~1,~1\}{]};

x2~=~Table{[}X2{[}i{]},~\{i,~0,~d~-~1,~1\}{]};

x3~=~Table{[}X3{[}i{]},~\{i,~0,~d~-~1,~1\}{]};
\end{lyxcode}
Defining the spin coordinates $S_{1I},S_{2I},S_{3I}$
\begin{lyxcode}
s1~=~Table{[}S1{[}i{]},~\{i,~1,~4,~1\}{]};

s2~=~Table{[}S2{[}i{]},~\{i,~1,~4,~1\}{]};

s3~=~Table{[}S3{[}i{]},~\{i,~1,~4,~1\}{]};
\end{lyxcode}
Defining the Dirac matrices
\begin{lyxcode}
G{[}0{]}~=~\{\{0,~1,~0,~0\},~\{-1,~0,~0,~0\},~\{0,~0,~0,~-1\},~\{0,~0,~1,~0\}\};

G{[}1{]}~=~\{\{0,~1,~0,~0\},~\{1,~0,~0,~0\},~\{0,~0,~0,~1\},~\{0,~0,~1,~0\}\};

G{[}2{]}~=~\{\{1,~0,~0,~0\},~\{0,~-1,~0,~0\},~\{0,~0,~1,~0\},~\{0,~0,~0,~-1\}\};

G{[}3{]}~=~\{\{0,~0,~0,~1\},~\{0,~0,~-1,~0\},~\{0,~0,~0,~0\},~\{0,~0,~0,~0\}\};

G{[}4{]}~=~\{\{0,~0,~0,~0\},~\{0,~0,~0,~0\},~\{0,~1,~0,~0\},~\{-1,~0,~0,~0\}\};

g~=~\{G{[}0{]},~G{[}1{]},~G{[}2{]},~G{[}3{]},~G{[}4{]}\};({*}These~are~$\Gamma_{A}${*})

gg~=~\{-G{[}0{]},~G{[}1{]},~G{[}2{]},~-2~G{[}4{]},~-2~G{[}3{]}\};({*}These~are~$\Gamma^{A}${*})
\end{lyxcode}
Defining the spacetime coordinates in spin space $X_{\,J}^{I}$
\begin{lyxcode}
X1IJ~=~Sum{[}X1{[}i{]}~G{[}i{]},~\{i,~0,~4\}{]};

X2IJ~=~Sum{[}X2{[}i{]}~G{[}i{]},~\{i,~0,~4\}{]};

X3IJ~=~Sum{[}X3{[}i{]}~G{[}i{]},~\{i,~0,~4\}{]};
\end{lyxcode}
Defining the spin space symplectic form $\Omega^{IJ}=\Omega_{IJ}$
\begin{lyxcode}
Om~=~\{\{0,~0,~1,~0\},~\{0,~0,~0,~1\},~\{-1,~0,~0,~0\},~\{0,~-1,~0,~0\}\};
\end{lyxcode}
Defining spin bilinears
\begin{lyxcode}
S1S2~=~s1.Om.s2;({*}$\left\langle S_{1}S_{2}\right\rangle ${*})

S3X1X2S3~=~s3.X1IJ.X2IJ.Om.s3;({*}$\left\langle S_{3}X_{1}X_{2}S_{3}\right\rangle ${*})

S1X2S3~=~s1.X2IJ.Om.s3;({*}$\left\langle S_{1}X_{2}S_{3}\right\rangle ${*})

S2X1S3~=~s2.X1IJ.Om.s3;({*}$\left\langle S_{2}X_{1}S_{3}\right\rangle ${*})

S1S3~=~s1.Om.s3;({*}$\left\langle S_{1}S_{3}\right\rangle ${*})

S2S3~=~s2.Om.s3;({*}$\left\langle S_{2}S_{3}\right\rangle ${*})

S1X3S2~=~s1.X3IJ.Om.s2;({*}$\left\langle S_{1}X_{3}S_{2}\right\rangle ${*})
\end{lyxcode}
Defining Light-cone ($X^{2}=0$) and transversality ($S_{I}X_{\,J}^{I}=0$)
constraints
\begin{lyxcode}
r1lc~=~\{X1{[}3{]}~->~1,~X1{[}4{]}~->~-X1{[}0{]}\textasciicircum 2~+~X1{[}1{]}\textasciicircum 2~+~X1{[}2{]}\textasciicircum 2\};

r1tr~=~Solve{[}(s1.X1IJ~/.~r1lc)~==~\{0,~0,~0,~0\},~\{S1{[}1{]},~S1{[}2{]}\}{]};

r2lc~=~\{X2{[}3{]}~->~1,~X2{[}4{]}~->~-X2{[}0{]}\textasciicircum 2~+~X2{[}1{]}\textasciicircum 2~+~X2{[}2{]}\textasciicircum 2\};

r2tr~=~Solve{[}(s2.X2IJ~/.~r2lc)~==~\{0,~0,~0,~0\},~\{S2{[}1{]},~S2{[}2{]}\}{]};

r3lc~=~\{X3{[}3{]}~->~1,~X3{[}4{]}~->~-X3{[}0{]}\textasciicircum 2~+~X3{[}1{]}\textasciicircum 2~+~X3{[}2{]}\textasciicircum 2\};

r3tr~=~Solve{[}(s3.X3IJ~/.~r3lc)~==~\{0,~0,~0,~0\},~\{S3{[}1{]},~S3{[}2{]}\}{]};

r~=~Join{[}r1lc,~r1tr{[}{[}1{]}{]},~r2lc,~r2tr{[}{[}1{]}{]},~r3lc,~r3tr{[}{[}1{]}{]}{]};
\end{lyxcode}
Defining 3-pt correlation functions
\begin{lyxcode}
X12~=~-2~x1.eta.x2;

X23~=~-2~x2.eta.x3;

X31~=~-2~x3.eta.x1;

\ensuremath{\phi}1\ensuremath{\phi}2O~=~~Simplify{[}(S3X1X2S3)\textasciicircum J/

~~~~~~~~~(X12\textasciicircum ((\textgreek{D}1+\textgreek{D}2-\textgreek{D}+J)/2)~X23\textasciicircum ((\textgreek{D}2+\textgreek{D}-\textgreek{D}1+J)/2)~X31\textasciicircum ((\textgreek{D}1+\textgreek{D}-\textgreek{D}2+J)/2)){]};

\textgreek{y}1\textgreek{y}2O1~=~Simplify{[}S1S2~(S3X1X2S3)\textasciicircum J/

~~~~~~~~~(X12\textasciicircum ((2\textgreek{Dy}-\textgreek{D}+J+1)/2)~X23\textasciicircum ((\textgreek{D}+J)/2)~X31\textasciicircum ((\textgreek{D}+J)/2)){]};

\textgreek{y}1\textgreek{y}2O2~=~Simplify{[}S2S3~S1S3~(S3X1X2S3)\textasciicircum (J-1)/

~~~~~~~~~(X12\textasciicircum ((2\textgreek{Dy}-\textgreek{D}+J-1)/2)~X23\textasciicircum ((\textgreek{D}+J)/2)~X31\textasciicircum ((\textgreek{D}+J)/2)){]};

\textgreek{y}1\textgreek{y}2O3~=~Simplify{[}S2X1S3~S1S3~(S3X1X2S3)\textasciicircum (J-1)/

~~~~~~~~~(X12\textasciicircum ((2\textgreek{Dy}-\textgreek{D}+J)/2)~X23\textasciicircum ((\textgreek{D}+J-1)/2)~X31\textasciicircum ((\textgreek{D}+J+1)/2))~+

~~~~~~~~~S1X2S3~S2S3~(S3X1X2S3)\textasciicircum (J-1)/

~~~~~~~~~(X12\textasciicircum ((2\textgreek{Dy}-\textgreek{D}+J)/2)~X23\textasciicircum ((\textgreek{D}+J+1)/2)~X31\textasciicircum ((\textgreek{D}+J-1)/2)){]};

\textgreek{y}1\textgreek{y}2O4~=~Simplify{[}S2X1S3~S1S3~(S3X1X2S3)\textasciicircum (J-1)/

~~~~~~~~~(X12\textasciicircum ((2\textgreek{Dy}-\textgreek{D}+J)/2)~X23\textasciicircum ((\textgreek{D}+J-1)/2)~X31\textasciicircum ((\textgreek{D}+J+1)/2))~-

~~~~~~~~~S1X2S3~S2S3~(S3X1X2S3)\textasciicircum (J-1)/

~~~~~~~~~(X12\textasciicircum ((2\textgreek{Dy}-\textgreek{D}+J)/2)~X23\textasciicircum ((\textgreek{D}+J+1)/2)~X31\textasciicircum ((\textgreek{D}+J-1)/2)){]};
\end{lyxcode}
Checking the Fierz identities
\begin{lyxcode}
F1~=~Simplify{[}S1X2S3~S2X1S3~+~S1S2~S3X1X2S3~+~S2S3~S1S3~X12~/.r{]}

>\textcompwordmark >\textcompwordmark >~0

F2~=~Simplify{[}S1X3S2~S3X1X2S3~-~S1S3~S2X1S3~X23~-~S2S3~S1X2S3~X13~/.r{]}

>\textcompwordmark >\textcompwordmark >~0
\end{lyxcode}
Test points
\begin{lyxcode}
p1~=~\{X1{[}0{]}~->~1,~X1{[}1{]}~->~0,~X1{[}2{]}~->~1,~X1{[}3{]}~->~1,~X1{[}4{]}~->~0,

~~~~~~X2{[}0{]}~->~0,X2{[}1{]}~->~1,~X2{[}2{]}~->~0,~X2{[}3{]}~->~1,X2{[}4{]}~->~1,

~~~~~~X3{[}0{]}~->~0,~X3{[}1{]}~->~0,X3{[}2{]}~->~1,~X3{[}3{]}~->~1,~X3{[}4{]}~->~1\};

pp~=~\{X1{[}0{]}~->~0,~X1{[}1{]}~->~0,~X1{[}2{]}~->~0,~X1{[}3{]}~->~1,~X1{[}4{]}~->~0,

~~~~~~X2{[}0{]}~->~0,X2{[}1{]}~->~1,~X2{[}2{]}~->~0,~X2{[}3{]}~->~1,~X2{[}4{]}~->~1\};

ppp~=~\{X1{[}0{]}~->~0,~X1{[}1{]}~->~0,~X1{[}2{]}~->~0,~X1{[}3{]}~->~1,~X1{[}4{]}~->~0\};

p~=~\{X1{[}0{]}~->~0,~X1{[}1{]}~->~0,~X1{[}2{]}~->~0,~X1{[}3{]}~->~1,~X1{[}4{]}~->~0,

~~~~~X2{[}0{]}~->~0,X2{[}1{]}~->~1,~X2{[}2{]}~->~0,~X2{[}3{]}~->~1,X2{[}4{]}~->~1,

~~~~~X3{[}0{]}~->~0,~X3{[}1{]}~->~1/2,X3{[}2{]}~->~Sqrt{[}3{]}/2,~X3{[}3{]}~->~1,X3{[}4{]}~->~1\};

~~~~({*}Three~equidistant~points~in~the~t=0~plane~in~3D~spacetime{*})
\end{lyxcode}
Defining 3D Gamma matrices, spinors and bispinors
\begin{lyxcode}
G3{[}0{]}~=~\{\{0,~-1\},~\{1,~0\}\};

G3{[}1{]}~=~\{\{0,~1\},~\{1,~0\}\};

G3{[}2{]}~=~\{\{1,~0\},~\{0,~-1\}\};

g3~=~\{G3{[}0{]},~G3{[}1{]},~G3{[}2{]}\};~({*}These~are~$\gamma_{\alpha}${*})

gg3~=~\{-G3{[}0{]},~G3{[}1{]},~G3{[}2{]}\};~({*}These~are~$\gamma^{\alpha}${*})

s13~=~\{S1{[}3{]},~S1{[}4{]}\};

s23~=~\{S2{[}3{]},~S2{[}4{]}\};

eab~=~\{\{0,~-1\},~\{1,~0\}\};

s2gs1~=~\{s23.G3{[}0{]}.eab.s13,~s23.G3{[}1{]}.eab.s13,~s23.G3{[}2{]}.eab.s13\};

s1gs2~=~\{s13.G3{[}0{]}.eab.s23,~s13.G3{[}1{]}.eab.s23,~s13.G3{[}2{]}.eab.s23\};
\end{lyxcode}
3D bilocal action: $Action=x_{12}^{2}\left\langle s_{1}\d_{1}\d_{s_{2}}\right\rangle \left\langle s_{2}\d_{2}\d_{s_{1}}\right\rangle $
\begin{lyxcode}
Action{[}x\_{]}~:=~~X12~Sum{[}(s13.(gg3{[}{[}m{]}{]})){[}{[}b{]}{]}~(s23.(gg3{[}{[}n{]}{]})){[}{[}a{]}{]}

~~~~~~~~~~~~~~~D{[}x~/.r,~s13{[}{[}b{]}{]},~s23{[}{[}a{]}{]},~x1{[}{[}m{]}{]},~x2{[}{[}n{]}{]}{]},

~~~~~~~~~~~~~~~\{a,~1,~2\},~\{b,~1,~2\},~\{m,~1,~3\},~\{n,~1,~3\}{]}

A1~=~Simplify{[}Action{[}\textgreek{y}1\textgreek{y}2O1~/.r{]}~-

~~~~~((3+J-\textgreek{D})(J-\textgreek{D})\textgreek{y}1\textgreek{y}2O1~-~4J(\textgreek{D}-1)\textgreek{y}1\textgreek{y}2O2)~/.r~/.pp~/.\textgreek{Dy}~->~1{]}

>\textcompwordmark >\textcompwordmark >~0~({*}This~shows~$S_{qa}\left\langle \psi\psi O\right\rangle ^{1}=\left(3+J-\Delta\right)\left(J-\Delta\right)\left\langle \psi\psi O\right\rangle ^{1}-4J\left(\Delta-1\right)\left\langle \psi\psi O\right\rangle ^{2}${*})

A2~=~Simplify{[}Action{[}\textgreek{y}1\textgreek{y}2O2~/.r{]}~+

~~~~~((3+J-\textgreek{D})(1+J-\textgreek{D})\textgreek{y}1\textgreek{y}2O1~+~(2+J-\textgreek{D})(1+J-\textgreek{D})\textgreek{y}1\textgreek{y}2O2)~/.r~/.pp~/.\textgreek{Dy}~->~1{]}

>\textcompwordmark >\textcompwordmark >~0~({*}This~shows~$S_{qa}\left\langle \psi\psi O\right\rangle ^{2}=-\left(3+J-\Delta\right)\left(1+J-\Delta\right)\left\langle \psi\psi O\right\rangle ^{1}-(2+J-\text{\textgreek{D}})(1+J-\text{\textgreek{D}})\left\langle \psi\psi O\right\rangle ^{2}${*})

A3~=~Factor{[}Simplify{[}Action{[}\textgreek{y}1\textgreek{y}2O3~/.~r{]}/\textgreek{y}1\textgreek{y}2O3~/.r~/.pp~/.\textgreek{Dy}~->~1{]}{]}

>\textcompwordmark >\textcompwordmark >~-(2~+~J~-~\textgreek{D})~(-1~+~J~+~\textgreek{D})

({*}This~shows~$S_{qa}\left\langle \psi\psi O\right\rangle ^{3}=-\left(2+J-\Delta\right)\left(-1+J+\Delta\right)\left\langle \psi\psi O\right\rangle ^{3}${*})

A4~=~Factor{[}Simplify{[}Action{[}\textgreek{y}1\textgreek{y}2O4~/.~r{]}/\textgreek{y}1\textgreek{y}2O4~/.r~/.pp~/.\textgreek{Dy}~->~1{]}{]}

>\textcompwordmark >\textcompwordmark >~(2~+~J~-~\textgreek{D})~(-1~+~J~+~\textgreek{D})

({*}This~shows~$S_{qa}\left\langle \psi\psi O\right\rangle ^{4}=\left(2+J-\Delta\right)\left(-1+J+\Delta\right)\left\langle \psi\psi O\right\rangle ^{4}${*})

\end{lyxcode}
Checking the action of $S_{emb}$ on the 3-pt. structures
\begin{lyxcode}
ActionEmb{[}x\_{]}~:=~Sum{[}-X12~(s1.(gg{[}{[}a{]}{]})){[}{[}j{]}{]}~(s2.(gg{[}{[}b{]}{]})){[}{[}i{]}{]}

~~~~~~~~~~~~~~~~~D{[}x,~s1{[}{[}i{]}{]},~s2{[}{[}j{]}{]},~x1{[}{[}a{]}{]},~x2{[}{[}b{]}{]}{]},

~~~~~~~~~~~~~~~~~\{a,~1,~d\},~\{b,~1,~d\},~\{i,~1,~4\},~\{j,~1,~4\}{]}

AE1~=~Simplify{[}ActionEmb{[}\textgreek{y}1\textgreek{y}2O1{]}~-

~~~~~~((3+J-\textgreek{D})(J-\textgreek{D})\textgreek{y}1\textgreek{y}2O1~-~4J(\textgreek{D}-1)\textgreek{y}1\textgreek{y}2O2)~/.r~/.p~/.\textgreek{Dy}~->~1{]}

>\textcompwordmark >\textcompwordmark >~0~({*}This~shows~$S_{emb}\left\langle \psi\psi O\right\rangle ^{1}=\left(3+J-\Delta\right)\left(J-\Delta\right)\left\langle \psi\psi O\right\rangle ^{1}-4J\left(\Delta-1\right)\left\langle \psi\psi O\right\rangle ^{2}${*})

AE2~=~Simplify{[}ActionEmb{[}\textgreek{y}1\textgreek{y}2O2{]}~-

~~~~~~((3+J-\textgreek{D})(1+J-\textgreek{D})\textgreek{y}1\textgreek{y}2O1~+~(2+J-\textgreek{D})(1+J-\textgreek{D})\textgreek{y}1\textgreek{y}2O2)~/.r~/.p~/.\textgreek{Dy}~->~1{]}

>\textcompwordmark >\textcompwordmark >~0~({*}This~shows~$S_{emb}\left\langle \psi\psi O\right\rangle ^{2}=\left(3+J-\Delta\right)\left(1+J-\Delta\right)\left\langle \psi\psi O\right\rangle ^{1}+(2+J-\text{\textgreek{D}})(1+J-\text{\textgreek{D}})\left\langle \psi\psi O\right\rangle ^{2}${*})

AE3~=~Factor{[}Simplify{[}ActionEmb{[}\textgreek{y}1\textgreek{y}2O3{]}/\textgreek{y}1\textgreek{y}2O3~/.r~/.p~/.\textgreek{Dy}~->~1{]}{]}

>\textcompwordmark >\textcompwordmark >~0~({*}This~shows~$S_{qa}\left\langle \psi\psi O\right\rangle ^{3}=0${*})

AE4~=~Factor{[}Simplify{[}ActionEmb{[}\textgreek{y}1\textgreek{y}2O4{]}/\textgreek{y}1\textgreek{y}2O4~/.r~/.p~/.\textgreek{Dy}~->~1{]}{]}

>\textcompwordmark >\textcompwordmark >~0~({*}This~shows~$S_{qa}\left\langle \psi\psi O\right\rangle ^{4}=0${*})

\end{lyxcode}
\pagebreak{}

\end{document}